%% file: main.tex
\documentclass[12pt,a4paper]{article}
\pdfoutput=1
\usepackage{ifthen} % for conditional statements
\newboolean{pdflatex}
\setboolean{pdflatex}{true} % False for eps figures 

\newboolean{articletitles}
\setboolean{articletitles}{true} % False removes titles in references

\newboolean{uprightparticles}
\setboolean{uprightparticles}{false} %True for upright particle symbols

\newboolean{inbibliography}
\setboolean{inbibliography}{false} %True once you enter the bibliography

\usepackage{longtable} % only for template; not usually to be used in PAPERs

\usepackage{subfigure}

\def\fO           {\ensuremath{\Pf_{0}(980)}\xspace}
\def\BsPhifO      {\decay{\Bs}{\phi \Pf_{0}(980)}\xspace}
\def\pipi                 {\ensuremath{\pi^+ \pi^-}\xspace}
\def\BsPhipipi    {\decay{\Bs}{\phi \pip \pim}\xspace}
\def\BdPhipipi    {\decay{\Bd}{\phi \pip \pim}\xspace}

\def\BsPhiPhi     {\decay{\Bs}{\phi \phi}\xspace}
\def\BsPhiRho     {\decay{\Bs}{\phi \rho}\xspace}

\newcommand{\ftwotw}{\ensuremath{f_2(1270)}\xspace}

\def\kk           {\ensuremath{K^+ K^-}\xspace}
\def\kkpipi               {\ensuremath{K^+ K^- \pi^+ \pi^-}\xspace}
\newcommand{\mpipi}{\ensuremath{m_{\pi\pi}}\xspace}

\newcommand{\RNum}[1]{\uppercase\expandafter{\romannumeral #1\relax}}

\input{preamble}

\begin{document}

%%%%%%%%%%%%%%%%%%%%%%%%%
%%%%% Title     %%%%%%%%%
%%%%%%%%%%%%%%%%%%%%%%%%%
\renewcommand{\thefootnote}{\fnsymbol{footnote}}
\setcounter{footnote}{1}

% %%%%%%% CHOOSE TITLE PAGE--------
%\onecolumn
% \input{title-LHCb-ANA}
%\input{title-LHCb-CONF}
\input{title-LHCb-PAPER}

%\twocolumn
% %%%%%%%%%%%%% ---------

\renewcommand{\thefootnote}{\arabic{footnote}}
\setcounter{footnote}{0}

%%%%%%%%%%%%%%%%%%%%%%%%%%%%%%%%
%%%%%  Table of Content   %%%%%%
%%%%%%%%%%%%%%%%%%%%%%%%%%%%%%%%
%%%% Uncomment next 2 lines if desired
%\tableofcontents
%\cleardoublepage

%%%%%%%%%%%%%%%%%%%%%%%%%
%%%%% Main text %%%%%%%%%
%%%%%%%%%%%%%%%%%%%%%%%%%

\pagestyle{plain} % restore page numbers for the main text
\setcounter{page}{1}
\pagenumbering{arabic}

%% Uncomment during review phase. 
%% Comment before a final submission.
%\linenumbers

% You can include short sections directly in the main tex file.
% However, for larger papers it is desirable to split the text into
% several semiautonomous files, which can be revised independently.
% This is especially useful when developing a document in
% collaboration with several people, since then different parts can be
% edited independently.  This type of file organization is shown here.
% 

\input{introduction}

\input{detector}

\input{selection}

\input{massfit}

\input{angular}

\input{results}

\input{systematics}

\input{conclusions}

% Do not include this in analysis note and conference reports
\input{acknowledgements}

\addcontentsline{toc}{section}{References}
\setboolean{inbibliography}{true}
\bibliographystyle{LHCb}
\bibliography{main,LHCb-PAPER,LHCb-CONF,LHCb-DP,LHCb-TDR,local}

%\input{main.bbl}

%\input{appendix}
% This should be taken out in the final paper
%\input{supplementary-app}

\newpage

% Author List ----------------------------                                                                                                                                                                                                                                                                                                
%  You need to get a new author list!                                                                                                                                                                                                                                                                                                    

%\input{LHCb_HD_authorlist_2014-06-20}
 
\newpage
\input{LHCb_Authorship_flat_29-Jun-2016.tex}

\end{document}

%% file: preamble.tex
\usepackage[top=1in, bottom=1.25in, left=1in, right=1in]{geometry}
\usepackage{microtype}
\usepackage{lineno}  % for line numbering during review
\usepackage{xspace} % To avoid problems with missing or double spaces after
\usepackage{caption} %these three command get the figure and table captions automatically small

\usepackage{graphicx}  % to include figures (can also use other packages)
\usepackage{color}
\usepackage{colortbl}
\graphicspath{{./figs/}} % Make Latex search fig subdir for figures

\usepackage{amsmath} % Adds a large collection of math symbols
\usepackage{amssymb}
\usepackage{amsfonts}
\usepackage{upgreek} % Adds in support for greek letters in roman typeset

\newcommand*\patchAmsMathEnvironmentForLineno[1]{%
\expandafter\let\csname old#1\expandafter\endcsname\csname #1\endcsname
\expandafter\let\csname oldend#1\expandafter\endcsname\csname
end#1\endcsname
 \renewenvironment{#1}%
   {\linenomath\csname old#1\endcsname}%
   {\csname oldend#1\endcsname\endlinenomath}%
}
\newcommand*\patchBothAmsMathEnvironmentsForLineno[1]{%
  \patchAmsMathEnvironmentForLineno{#1}%
  \patchAmsMathEnvironmentForLineno{#1*}%
}
\AtBeginDocument{%
\patchBothAmsMathEnvironmentsForLineno{equation}%
\patchBothAmsMathEnvironmentsForLineno{align}%
\patchBothAmsMathEnvironmentsForLineno{flalign}%
\patchBothAmsMathEnvironmentsForLineno{alignat}%
\patchBothAmsMathEnvironmentsForLineno{gather}%
\patchBothAmsMathEnvironmentsForLineno{multline}%
\patchBothAmsMathEnvironmentsForLineno{eqnarray}%
}

\usepackage{hyperref}    % Hyperlinks in references
\usepackage[all]{hypcap} % Internal hyperlinks to floats.

\input{lhcb-symbols-def} % Add in the predefined LHCb symbols

\usepackage{cite} % Allows for ranges in citations
\usepackage{mciteplus}

%% file: lhcb-symbols-def.tex
\def\lhcb {\mbox{LHCb}\xspace}

\def\MagUp {\mbox{\em Mag\kern -0.05em Up}\xspace}

\ifthenelse{\boolean{uprightparticles}}%
{

 \def\Ppi         {\ensuremath{\uppi}\xspace}

 \def\Ppsi        {\ensuremath{\uppsi}\xspace}

 \def\PDelta      {\ensuremath{\Delta}\xspace}                 
 \def\PXi      {\ensuremath{\Xi}\xspace}                 
 \def\PLambda      {\ensuremath{\Lambda}\xspace}                 
 \def\PSigma      {\ensuremath{\Sigma}\xspace}                 
 \def\POmega      {\ensuremath{\Omega}\xspace}                 
 \def\PUpsilon      {\ensuremath{\Upsilon}\xspace}                 
 
 \def\PB      {\ensuremath{\mathrm{B}}\xspace}                 
                  
 \def\PD      {\ensuremath{\mathrm{D}}\xspace}

 \def\PJ      {\ensuremath{\mathrm{J}}\xspace}                 
 \def\PK      {\ensuremath{\mathrm{K}}\xspace}

 \def\Pb      {\ensuremath{\mathrm{b}}\xspace}                 
 \def\Pc      {\ensuremath{\mathrm{c}}\xspace}

 \def\Pf      {\ensuremath{\mathrm{f}}\xspace}

 \def\Pi      {\ensuremath{\mathrm{i}}\xspace}

 \def\Ps      {\ensuremath{\mathrm{s}}\xspace}

}
{

 \def\Ppi         {\ensuremath{\pi}\xspace}

 \def\Ppsi        {\ensuremath{\psi}\xspace}                 
                  
 \mathchardef\PDelta="7101
 \mathchardef\PXi="7104
 \mathchardef\PLambda="7103
 \mathchardef\PSigma="7106
 \mathchardef\POmega="710A
 \mathchardef\PUpsilon="7107
                  
 \def\PB      {\ensuremath{B}\xspace}                 
                  
 \def\PD      {\ensuremath{D}\xspace}

 \def\PJ      {\ensuremath{J}\xspace}                 
 \def\PK      {\ensuremath{K}\xspace}

 \def\Pb      {\ensuremath{b}\xspace}                 
 \def\Pc      {\ensuremath{c}\xspace}

 \def\Pf      {\ensuremath{f}\xspace}

 \def\Pi      {\ensuremath{i}\xspace}

 \def\Ps      {\ensuremath{s}\xspace}

}

\makeatletter
\ifcase \@ptsize \relax% 10pt
  \newcommand{\miniscule}{\@setfontsize\miniscule{4}{5}}% \tiny: 5/6
\or% 11pt
  \newcommand{\miniscule}{\@setfontsize\miniscule{5}{6}}% \tiny: 6/7
\or% 12pt
  \newcommand{\miniscule}{\@setfontsize\miniscule{5}{6}}% \tiny: 6/7
\fi
\makeatother

\DeclareRobustCommand{\optbar}[1]{\shortstack{{\miniscule (\rule[.5ex]{1.25em}{.18mm})}
  \\ [-.7ex] $#1$}}

\def\squark    {{\ensuremath{\Ps}}\xspace}

\def\cquark    {{\ensuremath{\Pc}}\xspace}

\def\bquark    {{\ensuremath{\Pb}}\xspace}

\def\pion   {{\ensuremath{\Ppi}}\xspace}

\def\pip    {{\ensuremath{\pion^+}}\xspace}
\def\pim    {{\ensuremath{\pion^-}}\xspace}

\def\kaon    {{\ensuremath{\PK}}\xspace}
  \def\Kbar    {{\kern 0.2em\overline{\kern -0.2em \PK}{}}\xspace}

\def\KorKbar    {\kern 0.18em\optbar{\kern -0.18em K}{}\xspace}

\def\Kstar   {{\ensuremath{\kaon^*}}\xspace}

  \def\Dbar    {{\kern 0.2em\overline{\kern -0.2em \PD}{}}\xspace}
\def\D       {{\ensuremath{\PD}}\xspace}

\def\DorDbar    {\kern 0.18em\optbar{\kern -0.18em D}{}\xspace}
\def\Dz      {{\ensuremath{\D^0}}\xspace}

\def\Dm      {{\ensuremath{\D^-}}\xspace}

\def\Dsm     {{\ensuremath{\D^-_\squark}}\xspace}

\def\B       {{\ensuremath{\PB}}\xspace}
\def\Bbar    {{\ensuremath{\kern 0.18em\overline{\kern -0.18em \PB}{}}}\xspace}

\def\BorBbar    {\kern 0.18em\optbar{\kern -0.18em B}{}\xspace}
\def\Bz      {{\ensuremath{\B^0}}\xspace}

\def\Bd      {{\ensuremath{\B^0}}\xspace}
\def\Bs      {{\ensuremath{\B^0_\squark}}\xspace}
\def\Bsb     {{\ensuremath{\Bbar{}^0_\squark}}\xspace}

\def\jpsi     {{\ensuremath{{\PJ\mskip -3mu/\mskip -2mu\Ppsi\mskip 2mu}}}\xspace}

  \def\Y#1S{\ensuremath{\PUpsilon{(#1S)}}\xspace}% no space before {...}!

\def\Lz          {{\ensuremath{\PLambda}}\xspace}
\def\Lbar        {{\ensuremath{\kern 0.1em\overline{\kern -0.1em\PLambda}}}\xspace}
\def\LorLbar    {\kern 0.18em\optbar{\kern -0.18em \PLambda}{}\xspace}

\def\Lc      {{\ensuremath{\Lz^+_\cquark}}\xspace}

\newcommand{\decay}[2]{\ensuremath{#1\!\to #2}\xspace}         % {\Pa}{\Pb \Pc}

\def\to                 {\ensuremath{\rightarrow}\xspace}

\def\CP                {{\ensuremath{C\!P}}\xspace}

\def\AT#1     {\ensuremath{A_{\mathrm{T}}^{#1}}\xspace}           % 2

\def\C#1      {\ensuremath{\mathcal{C}_{#1}}\xspace}                       % 9
\def\Cp#1     {\ensuremath{\mathcal{C}_{#1}^{'}}\xspace}                    % 7
\def\Ceff#1   {\ensuremath{\mathcal{C}_{#1}^{\mathrm{(eff)}}}\xspace}        % 9  
\def\Cpeff#1  {\ensuremath{\mathcal{C}_{#1}^{'\mathrm{(eff)}}}\xspace}       % 7
\def\Ope#1    {\ensuremath{\mathcal{O}_{#1}}\xspace}                       % 2
\def\Opep#1   {\ensuremath{\mathcal{O}_{#1}^{'}}\xspace}                    % 7

\newcommand{\tev}{\ifthenelse{\boolean{inbibliography}}{\ensuremath{~T\kern -0.05em eV}\xspace}{\ensuremath{\mathrm{\,Te\kern -0.1em V}}}\xspace}
\newcommand{\gev}{\ensuremath{\mathrm{\,Ge\kern -0.1em V}}\xspace}
\newcommand{\mev}{\ensuremath{\mathrm{\,Me\kern -0.1em V}}\xspace}
\newcommand{\kev}{\ensuremath{\mathrm{\,ke\kern -0.1em V}}\xspace}
\newcommand{\ev}{\ensuremath{\mathrm{\,e\kern -0.1em V}}\xspace}
\newcommand{\gevc}{\ensuremath{{\mathrm{\,Ge\kern -0.1em V\!/}c}}\xspace}
\newcommand{\mevc}{\ensuremath{{\mathrm{\,Me\kern -0.1em V\!/}c}}\xspace}
\newcommand{\gevcc}{\ensuremath{{\mathrm{\,Ge\kern -0.1em V\!/}c^2}}\xspace}
\newcommand{\gevgevcccc}{\ensuremath{{\mathrm{\,Ge\kern -0.1em V^2\!/}c^4}}\xspace}
\newcommand{\mevcc}{\ensuremath{{\mathrm{\,Me\kern -0.1em V\!/}c^2}}\xspace}

\def\mum  {\ensuremath{{\,\upmu\rm m}}\xspace}

\newcommand{\chisqndf}{\ensuremath{\chi^2/\mathrm{ndf}}\xspace}

\def\gsim{{~\raise.15em\hbox{$>$}\kern-.85em
          \lower.35em\hbox{$\sim$}~}\xspace}
\def\lsim{{~\raise.15em\hbox{$<$}\kern-.85em
          \lower.35em\hbox{$\sim$}~}\xspace}

\newcommand{\Real}{\ensuremath{\mathcal{R}e}\xspace}

\def\sPlot{\mbox{\em sPlot}\xspace}
\def\pt         {\mbox{$p_{\rm T}$}\xspace}

\def\evtgen     {\mbox{\textsc{EvtGen}}\xspace}

\def\geant      {\mbox{\textsc{Geant4}}\xspace}

\def\photos     {\mbox{\textsc{Photos}}\xspace}

\def\pythia     {\mbox{\textsc{Pythia}}\xspace}

\def\tell1  {TELL1\xspace}
\def\ukl1   {UKL1\xspace}

%% file: title-LHCb-PAPER.tex
% $Id: title-LHCb-PAPER.tex 65671 2015-01-09 14:59:08Z tgershon $
% ===============================================================================
% Purpose: LHCb-PAPER journal paper title page template
% Author: 
% Created on: 2010-09-25
% ===============================================================================

%%%%%%%%%%%%%%%%%%%%%%%%%
%%%%%  TITLE PAGE  %%%%%%
%%%%%%%%%%%%%%%%%%%%%%%%%
\begin{titlepage}
\pagenumbering{roman}

% Header ---------------------------------------------------
\vspace*{-1.5cm}
\centerline{\large EUROPEAN ORGANIZATION FOR NUCLEAR RESEARCH (CERN)}
\vspace*{1.5cm}
\noindent
\begin{tabular*}{\linewidth}{lc@{\extracolsep{\fill}}r@{\extracolsep{0pt}}}
%\ifthenelse{\boolean{pdflatex}}% Logo format choice
%{\vspace*{-2.7cm}\mbox{\!\!\!\includegraphics[width=.14\textwidth]{lhcb-logo.pdf}} & &}%
%{\vspace*{-1.2cm}\mbox{\!\!\!\includegraphics[width=.12\textwidth]{lhcb-logo.eps}} & &}%
%\\
 & & CERN-EP-2016-213 \\  % ID 
 & & LHCb-PAPER-2016-028 \\  % ID 
 & & 23 November 2016 \\ 
 & & \\
\end{tabular*}

\vspace*{4.0cm}

% Title --------------------------------------------------
{\bf\boldmath\huge
\begin{center}   
  Observation of the decay $B^0_s\xspace\to\xspace\phi\pi^+\pi^-\xspace$ and evidence for 
  $B^0\xspace\to\xspace\phi\pi^+\pi^-\xspace$
\end{center}
}

\vspace*{1.0cm}

% Authors -------------------------------------------------
\begin{center}
%In the footnote, replace 'paper' by 'letter' in case of submission to PRL or PLB 
The LHCb collaboration\footnote{Authors are listed at the end of this paper.}
\end{center}

\vspace{\fill}

% Abstract -----------------------------------------------
\begin{abstract}
  \noindent The first observation of the rare decay $B^0_s\xspace\to\xspace\phi\pi^+\pi^-\xspace$
  and evidence for $B^0\xspace\to\xspace\phi\pi^+\pi^-\xspace$ are reported, using  
  $pp$ collision data recorded by the LHCb detector at centre-of-mass energies 
  $\sqrt{s} = 7$ and 8 TeV, corresponding to an integrated luminosity of 3~fb$^{-1}$. 
  The branching fractions in the $\pi^+\pi^-$ invariant mass range 
  $400<m(\pi^+\pi^-)<1600{\mathrm{\,Me\kern -0.1em V\!/}c^2}$ are 
  $[3.48\pm 0.23\pm 0.17\pm 0.35]\times 10^{-6}$ and
  $[1.82\pm 0.25\pm 0.41\pm 0.14]\times 10^{-7}$ for 
  $B^0_s\xspace\to\xspace\phi\pi^+\pi^-\xspace$
  and $B^0\xspace\to\xspace\phi\pi^+\pi^-\xspace$  respectively, 
  where the uncertainties are statistical, systematic and from the normalisation mode 
  $B^0_s\xspace\to\xspace\phi\phi\xspace$. A combined analysis of the $\pi^+\pi^-$ 
  mass spectrum and the decay angles of the final-state particles identifies
  the exclusive decays $\B^0_s\xspace\to\xspace\phi f_0(980)\xspace$, 
  $B_s^0\xspace\to\xspace\phi f_2(1270)\xspace$, and 
  $B^0_s\xspace\to\xspace\phi\rho^0\xspace$
  with branching fractions of 
  $[1.12\pm 0.16^{+0.09}_{-0.08}\pm 0.11]\times 10^{-6}$, 
  $[0.61\pm 0.13^{+0.12}_{-0.05}\pm 0.06]\times 10^{-6}$ and 
  $[2.7\pm 0.7\pm 0.2\pm 0.2]\times 10^{-7}$, respectively.
 
\end{abstract}

\vspace*{1.0cm}

\begin{center}
  Published on 11th January 2017 as Phys.Rev.D {\bf 95}, 026007 (2017)
\end{center}

\vspace{\fill}

{\footnotesize 
\centerline{\copyright~CERN on behalf of the \lhcb collaboration, licence \href{http://creativecommons.org/licenses/by/4.0/}{CC-BY-4.0}.}}
\vspace*{2mm}

\end{titlepage}

%%%%%%%%%%%%%%%%%%%%%%%%%%%%%%%%
%%%%%  EOD OF TITLE PAGE  %%%%%%
%%%%%%%%%%%%%%%%%%%%%%%%%%%%%%%%

%  empty page follows the title page ----
\newpage
\setcounter{page}{2}
\mbox{~}
%\newpage
%
%% Author List ----------------------------
%%  You need to get a new author list!
%\input{LHCb_authorlist.tex}
%
%The author list for journal publications is provided by the Membership Committee shortly after 'approval to go to paper' has been given.
%%It will be made available on the page
%%\verb!http://www.physik.uzh.ch/~strauman/forMemCo/LHCb-PAPER-XXXX-XXX/! .
%It will be sent to you by email shortly after a paper number has beens assigned.
%The author list should be included already at first circulation, 
%to allow new members of the collaboration to verify whether they have been included correctly.
%Occasionally a misspelled name is corrected or associated institutions become full members.
%In that case, a new author list will be sent to you.
%In case line numbering doesn't work well after including the authorlist, try moving the \verb!\bigskip! after the last author to a separate line.
%
%
%The authorship for Conference Reports should be ``The LHCb
%  collaboration'', with a footnote giving the name(s) of the contact
%  author(s), but without the full list of collaboration names.

\cleardoublepage

%% file: introduction.tex
\section{Introduction}
\label{sec:Introduction}
\begin{figure}[tb]
\begin{center}
\subfigure[]{
\includegraphics[scale=1.2]{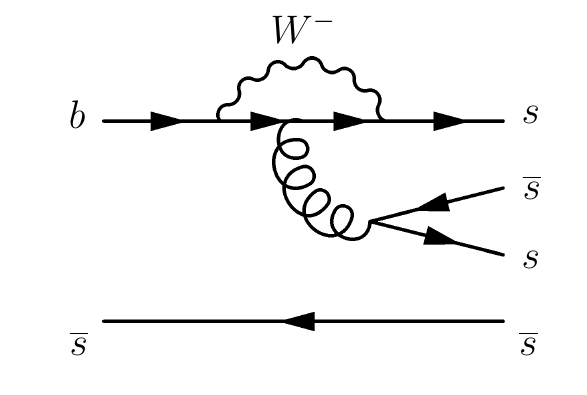}
}
\subfigure[]{
\includegraphics[scale=1.2]{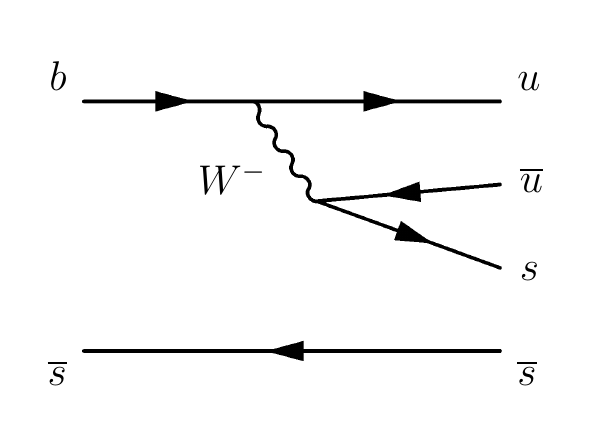}
}
\caption{\small Feynman diagrams for the exclusive decays (a) $B_s^0 \to \phi \fO$ 
and (b) $B_s^0 \to \phi \rho^0$.}
\label{fig:feynman}
\end{center}
\end{figure}

The decays $\BsPhipipi$ and $\BdPhipipi$  have not been observed before. They are 
examples of decays that are dominated by contributions from 
flavour changing neutral currents (FCNC), which provide a sensitive probe for
the effect of physics beyond the Standard Model because their  
amplitudes are described by loop (or penguin) 
diagrams where new particles may enter~\cite{Raidal:2002ph}. 
A well-known example of this type of decay is $\BsPhiPhi$ which has a 
branching fraction of $1.9\times 10^{-5}$~\cite{PDG2014}. 
First measurements of the \CP-violating phase
$\phi_s$ in this mode have recently been made by the LHCb
collaboration~\cite{LHCb-PAPER-2013-007,LHCb-PAPER-2014-026}. 
The decay \BsPhifO also proceeds via a gluonic $b\to s$ penguin transition (see Fig.~\ref{fig:feynman}(a)), with 
an expected branching fraction of approximately $2\times 10^{-6}$, based on the ratio of the  
$\Bs\to J/\psi f_0(980)$ and $\Bs\to J/\psi \phi$ decays~\cite{PDG2014}. 
When large statistics samples are available, similar time-dependent \CP violation studies 
will be possible with $\BsPhifO$.

The decay $\Bs\rightarrow \phi\rho^0$ is of particular 
interest\footnote{Unless otherwise stated, $\rho^0$ represents the 
$\rho(770)^0$, $\Kstar^0$ represents the $K^*(892)^0$, 
$\phi$ represents the $\phi(1020)$, 
and charge-conjugate decays are implied throughout this paper.}, 
because it is an isospin-violating $\Delta I = 1$ transition which is mediated by 
a combination of an electroweak penguin diagram and a 
suppressed $b\to u$ transition (see Fig.~\ref{fig:feynman}(b)). 
The predicted branching fraction is $[4.4^{+2.2}_{-0.7}]\times 10^{-7}$,
and large \CP-violating asymmetries are not excluded~\cite{Hofer:2010ee}.  

The corresponding $\Bz$ decays are mediated by CKM-suppressed $b\to d$ penguin diagrams, 
and are expected to have branching fractions an order of magnitude lower than the $\Bs$ decays. 
The BaBar experiment has set an upper limit on the branching fraction 
of the decay $\Bz\rightarrow\phi\rho^0$ of $3.3 \times 10^{-7}$ 
at $90\%$ confidence level \cite{Aubert:2008fq}. 

This paper reports a time-integrated and flavour-untagged search, 
using a dataset with an integrated luminosity of approximately 3~fb$^{-1}$ 
collected by the LHCb detector in 2011 and 2012 at centre-of-mass energies of 
$\sqrt{s}=$7 and 8~TeV, respectively.
This leads to the first observation of $\BsPhipipi$ decays, 
and evidence for $\BdPhipipi$ decays, with the $\pipi$ invariant mass 
in the range $400<m(\pipi)<1600$\mevcc. 
A combined angular and $\pipi$ mass analysis of the $\BsPhipipi$ sample identifies
contributions from the exclusive decays $\BsPhifO$, $B_s^0 \to \phi \ftwotw$, 
and $\BsPhiRho^0$.  
There is also a significant S-wave $\pipi$ contribution in the high-mass region 
$1350<m(\pipi)<1600$\mevcc.  

The branching fractions for both the inclusive and exclusive decays  are determined 
with respect to the normalisation mode $\BsPhiPhi$. This mode has a very similar topology 
and a larger branching fraction, 
which has been measured by the LHCb collaboration~\cite{LHCb-PAPER-2015-028}
to be $\mathcal{B}(\BsPhiPhi) = [1.84\pm 0.05\pm 0.07\pm 0.11\pm 0.12]\times 10^{-5}$,
where the uncertainties are respectively statistical, systematic, 
from the fragmentation function $f_s/f_d$ giving the ratio of $\Bs$ to $\Bz$ production at the LHC, 
and from the measurement of the branching fraction of $\Bz\to\phi\Kstar^0$ 
at the \B factories~\cite{Aubert:2008zza, Prim:2013nmy}.

%% file: detector.tex
\section{Detector and software}
\label{sec:Detector}
The \lhcb detector~\cite{Alves:2008zz,LHCb-DP-2014-002} is a single-arm forward
spectrometer covering the \mbox{pseudorapidity} range $2<\eta <5$. 
It is designed for the study of particles containing \bquark or \cquark quarks, 
which are produced preferentially as pairs at small angles with respect to the beam axis. 
The detector includes a high-precision tracking system
consisting of a silicon-strip vertex detector surrounding the $pp$
interaction region,
a large-area silicon-strip tracker 
located upstream of a dipole magnet with a bending power of about
$4{\rm\,Tm}$, and three stations of silicon-strip trackers and straw
drift tubes placed downstream of the magnet.
The tracking system provides a measurement of charged particle momenta with
a relative uncertainty that varies from 0.5\% at 5\gevc to
1.0\% at 200\gevc. The minimum distance of a track to a primary $pp$ interaction 
vertex (PV), the impact parameter (IP), is measured with a resolution of $(15+29/\pt)\mum$,
where \pt is the component of the momentum transverse to the beam, in\,\gevc.
Different types of charged hadrons are distinguished using information
from two ring-imaging Cherenkov detectors.
Photon, electron and hadron candidates are identified by a calorimeter system consisting of
scintillating-pad and preshower detectors, an electromagnetic
calorimeter and a hadronic calorimeter. Muons are identified by a
system composed of alternating layers of iron and multiwire
proportional chambers.

The trigger~\cite{LHCb-DP-2012-004}  
consists of a hardware stage, based on information from the calorimeter and muon
systems, followed by a software stage, which applies a full event
reconstruction. The software trigger requires a two-, three- or four-track
  secondary vertex with a significant displacement from an associated PV. 
  At least one charged particle
  must have a transverse momentum $\pt > 1.7\gevc$ and be
  inconsistent with originating from the PV. 
  A multivariate algorithm~\cite{BBDT} is used for
  the identification of secondary vertices consistent with the decay
  of a \bquark hadron into charged hadrons. 
  In addition, an algorithm is used that identifies
  inclusive $\phi\to K^+K^-$ production at a secondary vertex, without requiring 
  a decay consistent with a \bquark hadron. 

In the simulation, $pp$ collisions are generated using
\pythia 6~\cite{Sjostrand:2006za}  with a specific \lhcb
configuration~\cite{LHCb-PROC-2010-056}.  Decays of hadronic particles
are described by \evtgen~\cite{Lange:2001uf}, in which final-state
radiation is generated using \photos~\cite{Golonka:2005pn}. The
interaction of the generated particles with the detector and its
response are implemented using the \geant
toolkit~\cite{Allison:2006ve, *Agostinelli:2002hh} as described in
Ref.~\cite{LHCb-PROC-2011-006}.

%% file: selection.tex
\section{Selection}
\label{sec:selection}
The offline selection of candidates consists of two parts. First, a
selection with loose criteria is performed that reduces the combinatorial background 
as well as removing some specific backgrounds from other
exclusive $\bquark$-hadron decay modes. In a second stage a multivariate method
is applied to further reduce the combinatorial background and 
improve the signal significance.

The selection starts from well-reconstructed particles
that traverse the entire spectrometer and have 
$p_T>500 \mevc$. Spurious tracks created by the 
reconstruction are suppressed using a neural
network trained to discriminate between these and real particles. 
A large track IP with respect to any PV is required, 
consistent with the track coming from a displaced secondary decay vertex.  
The information provided by the ring-imaging
Cherenkov detectors is combined with information 
from the tracking system to select charged
particles consistent with being a kaon, pion or proton. 
Tracks that are identified as muons are removed at this stage.
 
Pairs of oppositely charged kaons that originate from a common vertex
are combined to form a $\phi$ meson candidate. The transverse momentum of the
$\phi$ meson is required to be larger than $ 0.9 \gevc $ and the
invariant mass to be within $10 \mevcc$ of the known value
\cite{PDG2014}. Similarly, pairs of oppositely charged pions are
combined if they form a common vertex and if the transverse momentum
of the $\pipi$ system is larger than $1 \gevc$. For this analysis, the invariant mass of the pion pair 
is required to be in the range $400<m(\pipi)<1600 \mevcc$, below the charm threshold.     
The $\phi$ candidates and $\pipi$ pairs are combined to form $\Bd$ or $\Bs$
meson candidates. To further reject combinatorial background, 
the reconstructed flight path of the $B$ candidates must be consistent 
with coming from a PV.

There are several decays 
of $b$ hadrons proceeding via charmed hadrons that need to be 
explicitly removed. The decay modes $\Bs \rightarrow \Dsm \pi^+$ and $\Bz \rightarrow \Dm \pi^+$ 
are rejected when the invariant mass of the $K^+K^-\pi^-$ system is  
within 3 standard deviations ($\sigma$) of either $D$ meson mass. 
The decay mode $\BorBbar \rightarrow \DorDbar K^{\pm}\pi^{\mp}$ 
is rejected when the invariant mass of either of the $K^{\pm}\pi^{\mp}$ combinations
is within 2$\sigma$ of the $\Dz$ mass. 
Backgrounds from $D^-$ decays to $K^+\pi^-\pi^-$ and from $\Lc$ decays to $pK^-\pi^+$ 
are removed if the three-body invariant mass, calculated assuming that either a $\pi^-$ 
or a proton has been misidentified as a kaon, is within 3$\sigma$ of the charm hadron mass. 

Another background arises from the decay $\Bz \to \phi \Kstar^0$, where 
the kaon from the decay $\Kstar^0\to K^+\pi^-$ is misidentified as a pion. 
To remove it, the invariant masses $m(K^+ \pi^-)$ and $m(K^+ K^- K^+ \pi^-)$ are calculated
assuming that one of the $K^+$ has been misidentified as a $\pi^+$, and candidates are rejected if 
$m(K^+ \pi^-)$ is within 3 decay widths of the $\Kstar^0$, 
and $m(K^+ K^- K^+ \pi^-)$ is consistent with the $\Bz$ mass to 
within 3 times the experimental resolution.   
The higher resonance mode $\Bz \to \phi K_2^{*0}(1430)$ is vetoed in a similar fashion. 
The efficiency of the charm and $\phi\Kstar^0$ vetoes is 94\%,
evaluated on the $\BsPhipipi$ simulation sample,
with the $\phi\Kstar^0$ veto being 99\% efficient. 
For the decay $\BdPhipipi$ this efficiency is reduced to 84\% 
by the larger impact of the $\phi\Kstar^0$ veto.

In the second stage of the selection a boosted decision tree~(BDT)~\cite{Breiman,AdaBoost} 
is employed to further reduce the combinatorial background. This makes use of twelve
variables related to the kinematics of the $B$ meson
candidate and its decay products, particle identification for the kaon candidates and 
the $B$ decay vertex displacement from the PV. 
It is trained using half of both the simulated signal sample and the background events 
from the data in the range $5450 < m(K^+ K^-\pi^+ \pi^-) < 5600 \mevcc$, 
and validated using the other half of each sample.
For a signal efficiency of 90\% the BDT has a background rejection of 99\%. 

A sample of \BsPhiPhi candidates has been selected using 
the same methods as for the signal modes, apart from the particle identification criteria 
and the $m(K^+K^-)$ mass window for the second $\phi$ meson, and without the 
$\phi K^{*0}$ veto. 
The BDT deliberately does not include particle identification 
for the pion candidates, because this part of the selection is different between the signal mode and 
the \BsPhiPhi normalisation mode. 

For the signal mode a tighter selection is made on the pion identification 
as part of a two-dimensional optimisation together with the BDT output. 
The figure of merit (FOM) used to maximise the discovery potential 
for a new signal is\cite{Punzi:2003bu},
\begin{equation}
{\rm FOM} = \frac{\varepsilon_{S}}{5/2+\sqrt{B}} \hspace{0.1cm},\nonumber 
\label{eq:phiphiFOM} 
\end{equation} 
where $\varepsilon_{S}$ is the signal efficiency evaluated using the
simulation and $B$ is the number of background candidates expected within a 50\mevcc window 
about the \Bs mass. The optimised selection on the BDT output and the pion identification has a signal 
efficiency $\varepsilon_{S} = 0.846$.

%% file: massfit.tex
\section{Invariant mass fit}
\label{sec:massfit}
The yields for the inclusive $\BsPhipipi$ and $\BdPhipipi$  signals are determined from 
a fit to the invariant \kkpipi mass distribution of selected candidates
in the range $5100<m(\kkpipi)<5600\mevcc$. 
The fit includes possible signal contributions from both $\Bs$ and $\Bz$ decays, 
as well as combinatorial background. 
Backgrounds from partially reconstructed decays such as $\Bs\to\phi\phi(\to\pi^+\pi^-\pi^0)$ 
and $\Bs\to\phi\eta'(\to\pi^+\pi^-\gamma)$ are negligible
in the region $m(\kkpipi)>5100\mevcc$. After the veto the contribution from 
$\Bz\to\phi K^{*0}$ can also be neglected.

The line shapes for the $\BsPhipipi$ signal and $\BsPhiPhi$ normalisation mode 
are determined using simulated events, and parameterised by a sum of two Gaussian functions
with a common mean and different widths. 
In the fits to data the means and widths of the narrow Gaussians 
for the $\Bs$ modes are fitted, but the relative widths and fractions of the broader Gaussians 
relative to the narrow ones are taken from the simulation. 
The mean and width of the $\Bz$ signal shape are scaled down 
from $\BsPhipipi$ to account for the mass difference~\cite{PDG2014}, and to correct for 
a slight modification of the $\Bz$ shape due to the $\phi\Kstar^0$ veto.
The combinatorial background is modelled by an exponential function with a slope 
that is a free parameter in the fit to the data. 

\begin{figure}[tb]
\centering
\includegraphics[height=11.cm]{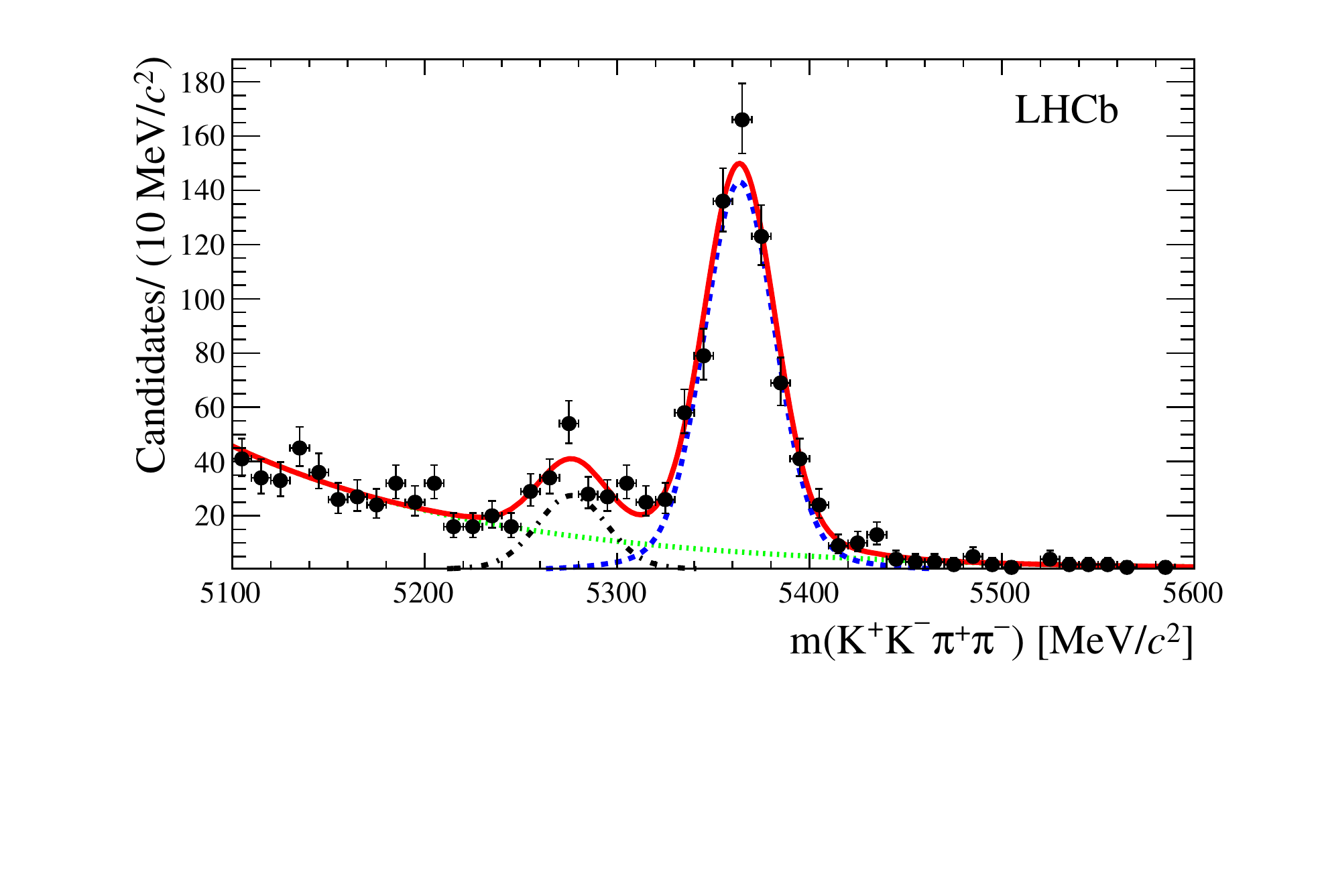}
\vspace{-3.5cm}
\caption{\small The \kkpipi invariant mass distribution for candidates in the
mass range $0.4 < \mpipi < 1.6$\gevcc. The fit described in the text
is overlaid. 
The solid (red) line is the total fitted function, the
dotted (green) line the combinatorial background, the dashed (blue) line
the $\Bs$ and the dot-dashed (black) line the $\Bz$ signal component.} 
\label{fig:massfit2}
\end{figure}

\begin{figure}[tb]
\centering
\includegraphics[height=14cm]{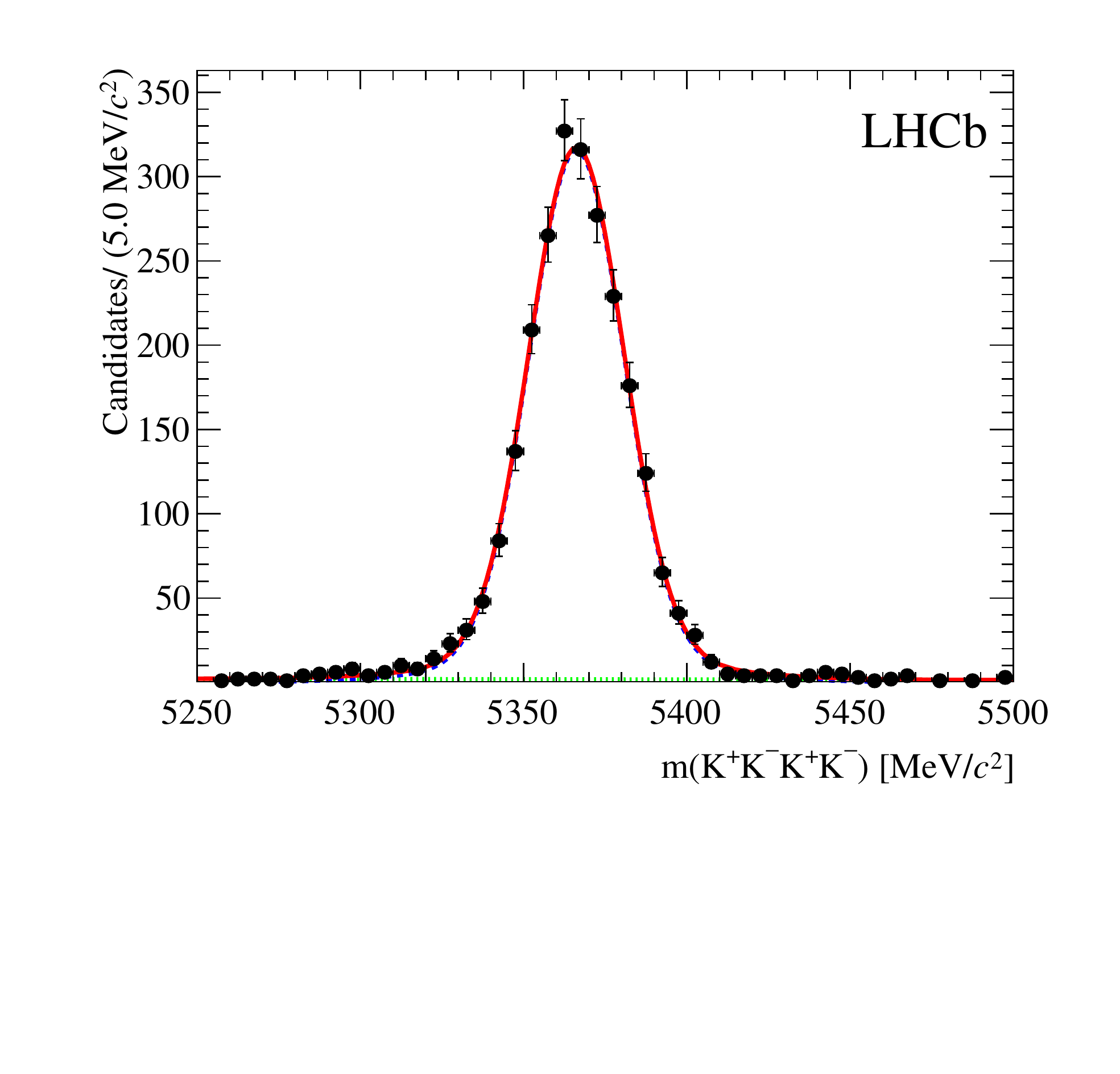}
\vspace{-4cm}
\caption{\small The $K^+ K^- K^+ K^-$ invariant mass distribution after all
  selection criteria. The solid (red) line is the total fitted function
  including the \BsPhiPhi signal, 
  and the dashed (green) line is the combinatorial background.}
\label{fig:phiphi_mass}
\end{figure}

Figure~\ref{fig:massfit2} shows the result of the extended unbinned maximum likelihood fit to 
the $m(\kkpipi)$ distribution. 
There is clear evidence for both $\BsPhipipi$ and $\BdPhipipi$ signals.
The $\Bs$ and $\Bz$ yields are $697\pm 30$ and $131\pm 17$ events, respectively,
and the fit has a chi-squared per degree of freedom, $\chisqndf$, of 0.87.
Figure~\ref{fig:phiphi_mass}  
shows the $m(K^+ K^- K^+ K^-)$ distribution for the \BsPhiPhi normalisation mode,
with a fit using a sum of two Gaussians for the \Bs signal shape. 
There are $2424\pm 51$ events above a very low combinatorial background. 
Backgrounds from other decay modes are negligible with this selection.

\begin{figure}[tb]
\centering
\includegraphics[width=0.8\linewidth]{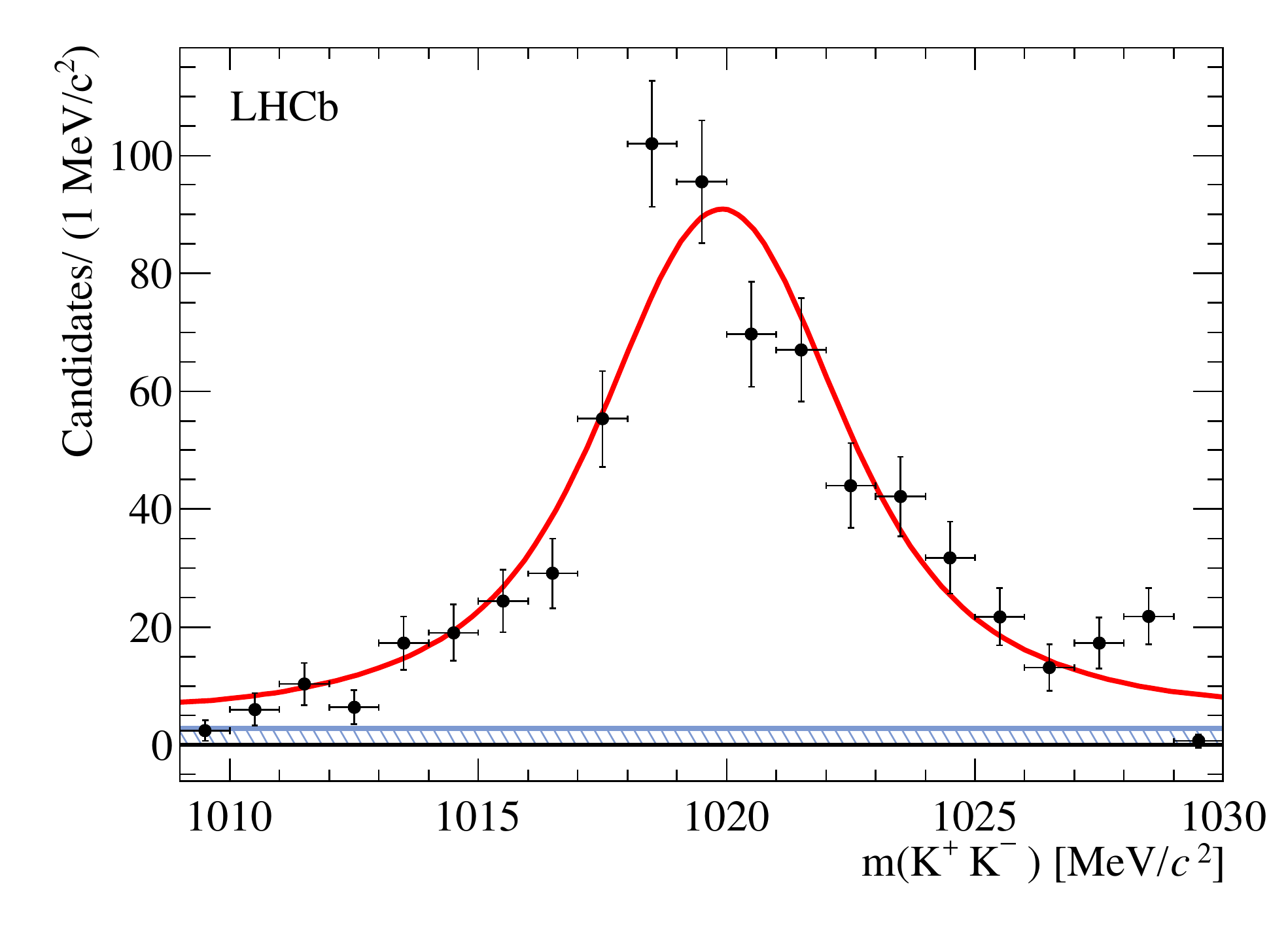}
\caption{\small The $\kk$ invariant mass distribution for background-subtracted 
$\BsPhipipi$ signal events with a fit to the dominant P-wave $\phi$ meson shown as a 
solid (red) line, and a small S-wave $\kk$ contribution shown as a hatched (blue) area.} 
\label{fig:resonance}
\end{figure}

To study the properties of the $\BsPhipipi$ signal events, the combinatorial background 
and $\Bz$ contribution are subtracted using the \sPlot method \cite{Pivk:2004ty}. 
The results of the invariant mass fit are used to assign 
to each event a signal weight that factorizes out the signal part of the sample from the
other contributions. These weights can then be used to project out other kinematic properties 
of the signal, provided that these properties are uncorrelated with $m(\kkpipi)$.
In the next section the decay angle and $m(\pipi)$ distributions of the 
$\BsPhipipi$ signal events are used to study the resonant $\pipi$ contributions. 
Figure~\ref{fig:resonance} shows the $\kk$ invariant mass distribution for the 
$\BsPhipipi$ signal, which is consistent
with a dominant $\phi$ meson resonance together with a small contribution
from a non-resonant S-wave \kk component. 
The $\phi$ contribution is modelled by a relativistic Breit-Wigner function, whose natural width is 
convolved with the experimental \kk mass resolution, and the S-wave component is modelled 
by a linear function. The S-wave \kk component is fitted to be $(8.5\pm 3.8)$\% 
of the signal yield in a $\pm 10\mevcc$ window around the known $\phi$ mass. 
A similar fit to the $\BsPhiPhi$ normalisation mode gives an S-wave component 
of $(1.4\pm 1.1)$\%. 

%% file: angular.tex
\section{Amplitude Analysis}
\label{sec:analysis}

There are several resonances that can decay into a $\pipi$ final state in the region 
$400<m(\pipi)<1600$\mevcc. These are listed in Table~\ref{tab:resonances} together 
with the mass models used to describe them and the source of the model parameters.\footnote{Note that 
the description of 
the broad $f_0(1370)$ and $f_0(1500)$ resonances by Breit-Wigner functions is known not to be a good 
approximation when they both make significant contributions~\cite{Bugg:2008eb}.}
To study the resonant contributions, an amplitude analysis 
is performed using an unbinned maximum likelihood fit to the $m(\pipi)$ mass and decay 
angle distributions of the \Bs candidates with their signal weights obtained by the \sPlot technique. 
In the fit the uncertainties on the signal weights are taken into account in determining 
the uncertainties on the fitted amplitudes and phases.
\begin{table}[tb]
\begin{center}
\caption{\small Possible resonances contributing to the $m(\pipi)$ mass distribution.
The shapes are either relativistic Breit-Wigner (BW) functions, or empirical 
threshold functions for the $f_0(500)$ proposed by Bugg~\cite{Bugg:2006gc} based on data from BES, 
and for the $f_0(980)$ proposed by Flatt\'e~\cite{Flatte:1976xv} to account for the effect of the 
$\kk$ threshold.}
\begin{tabular}{cccccc}
Resonance & Spin &  Shape & Mass & Width & Source \\
\hline 
$f_0(500)$     &  0  &  Bugg  & 400--800 & Broad & BES~\cite{Bugg:2006gc} \\    
$\rho$     &  1  &  BW   & 775 & 149 & PDG~\cite{PDG2014} \\
$f_0(980)$     &  0  &  Flatt\'e & 980 & 40--100 & LHCb~\cite{LHCb-PAPER-2012-005} \\
$f_2(1270)$     &  2  &  BW & 1275 & 185 & PDG~\cite{PDG2014} \\ 
$f_0(1370)$     &  0  &  BW & 1200--1500 & 200--500 & PDG~\cite{PDG2014} \\ 
$f_2(1430)$     &  2  &  BW & 1421 & 30 & DM2~\cite{Augustin:1987da} \\ 
$\rho(1450)$    &  1  &  BW & 1465 & 400 & PDG~\cite{PDG2014} \\
$f_0(1500)$     &  0  &  BW &  1461 & 124 & LHCb~\cite{LHCb-PAPER-2013-069}\\ \hline
\end{tabular}
\label{tab:resonances}
\end{center}
\end{table}

\begin{figure}[tb]
\centering
\includegraphics[height=6.cm]{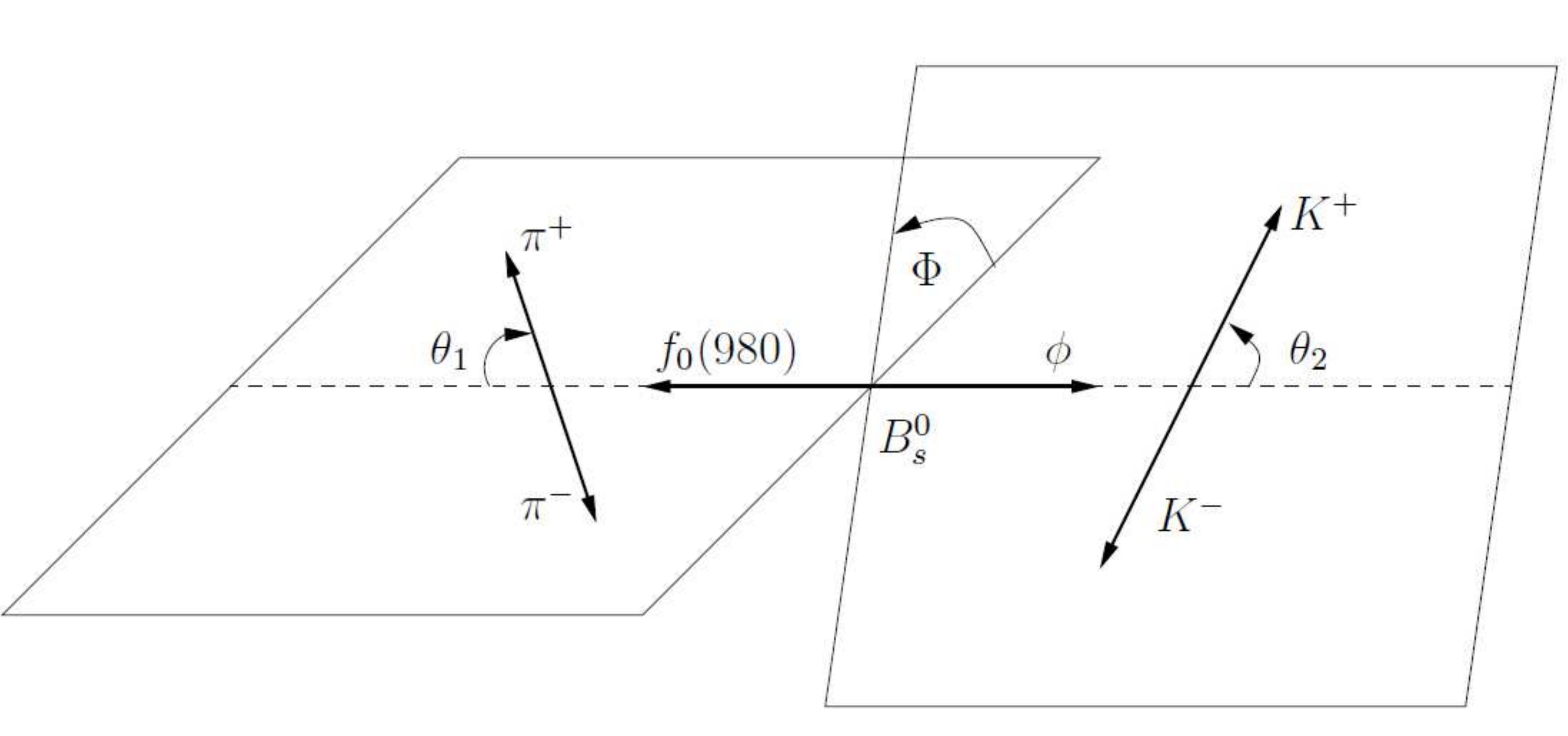}
\caption{The definition of the decay angles $\theta_1$, $\theta_2$ and $\Phi$ for the decay 
$\BsPhipipi$ with $\phi\to K^+K^-$ and taking $f_0(980)\to\pi^+\pi^-$ for illustration.} 
\label{fig:decayangles}
\end{figure}

Three decay angles are defined in the transversity basis as illustrated 
in Fig.~\ref{fig:decayangles}, 
where $\theta_1$ is the \pipi helicity angle between the $\pi^+$ direction in the 
$\pipi$ rest frame and the $\pipi$ direction in the $B$ rest frame,
$\theta_2$ is the $K^+K^-$ helicity angle between the $K^+$ direction in the $\phi$ rest frame 
and the $\phi$ direction in the $B$ rest frame, and $\Phi$ is the acoplanarity angle between
the $\pipi$ system and the $\phi$ meson decay planes.

\begin{figure}[tb]
\centering
\includegraphics[width=0.45\textwidth]{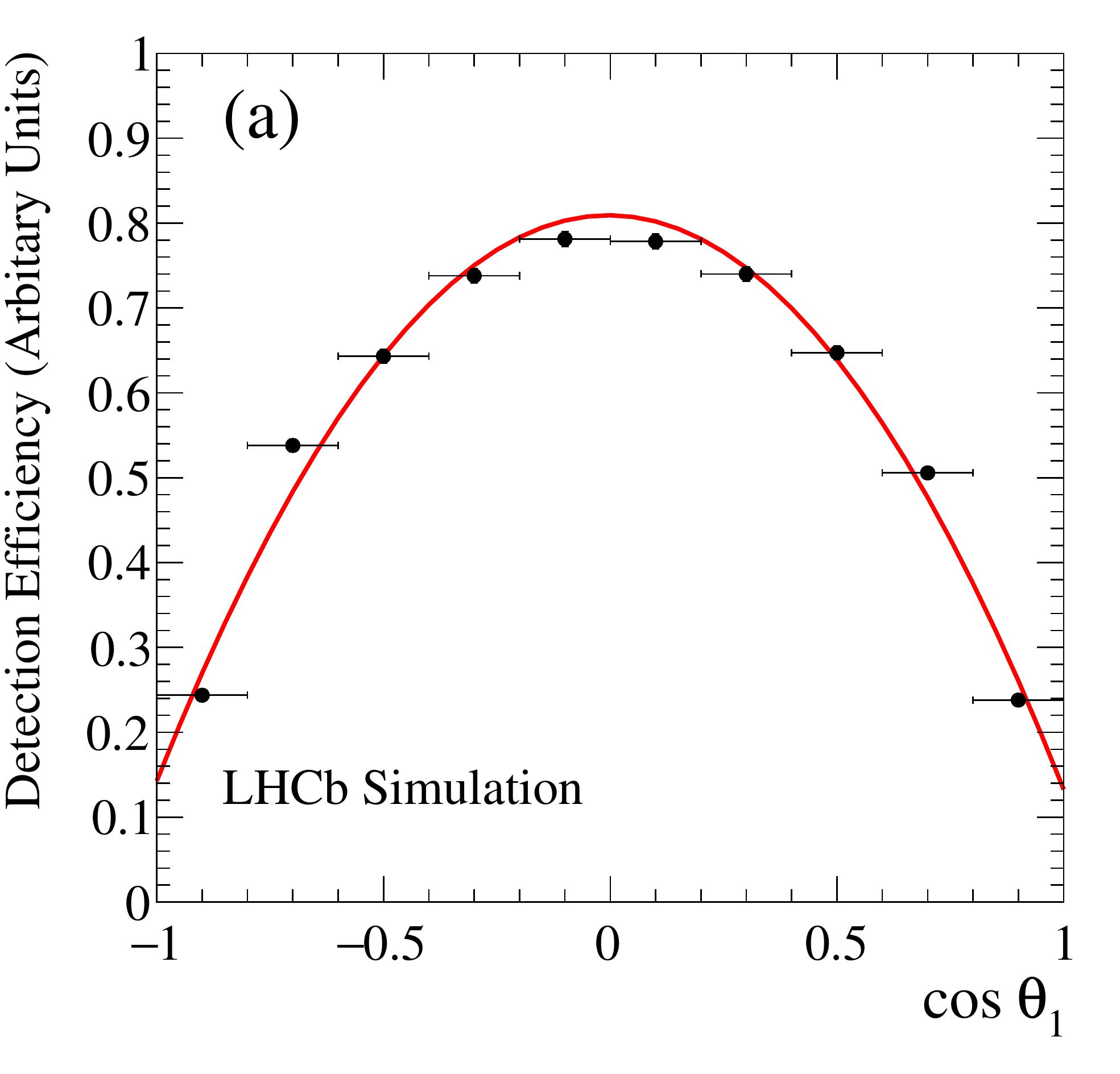} \hfill
\includegraphics[width=0.45\textwidth]{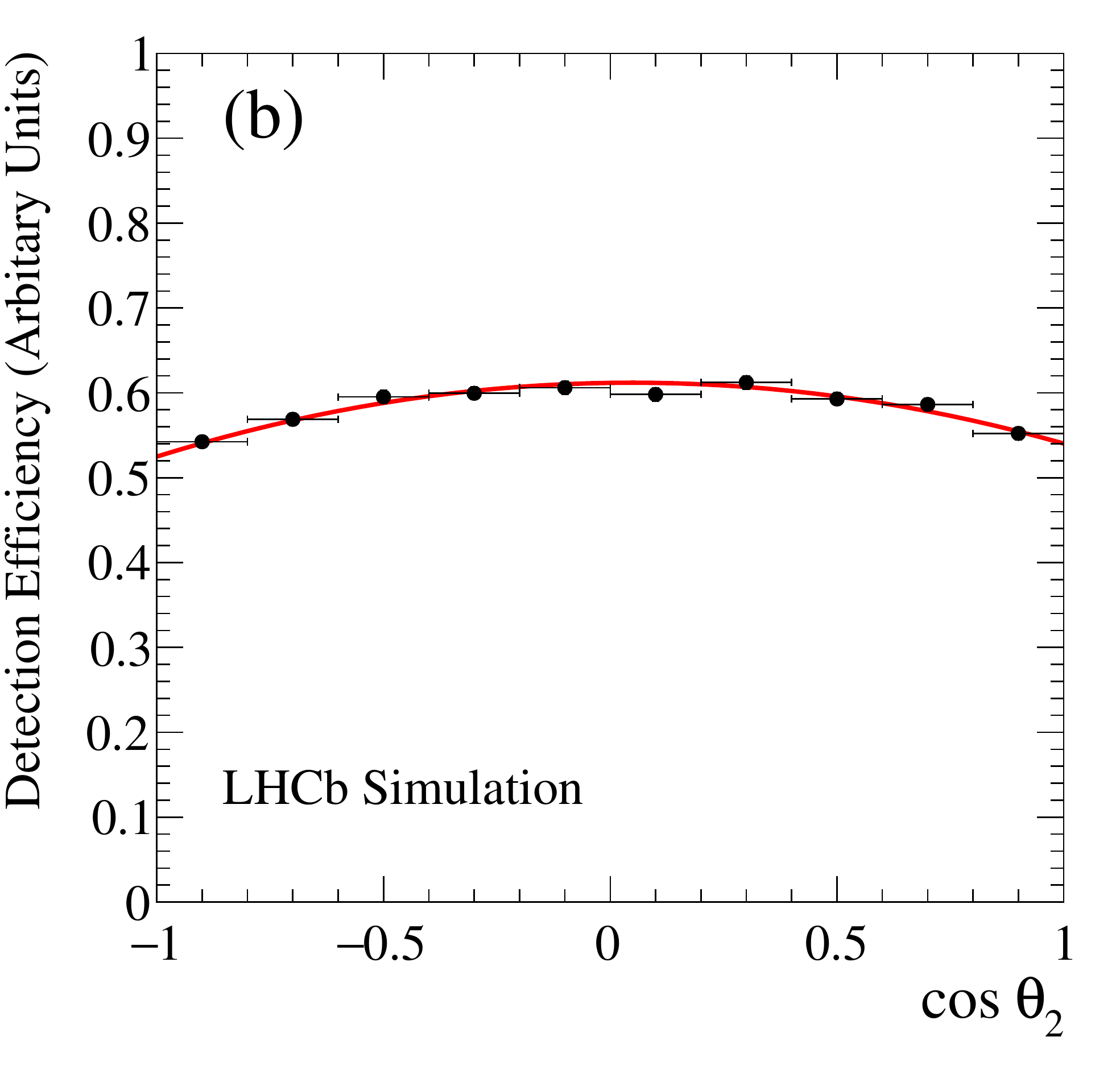} \\
\includegraphics[width=0.45\textwidth]{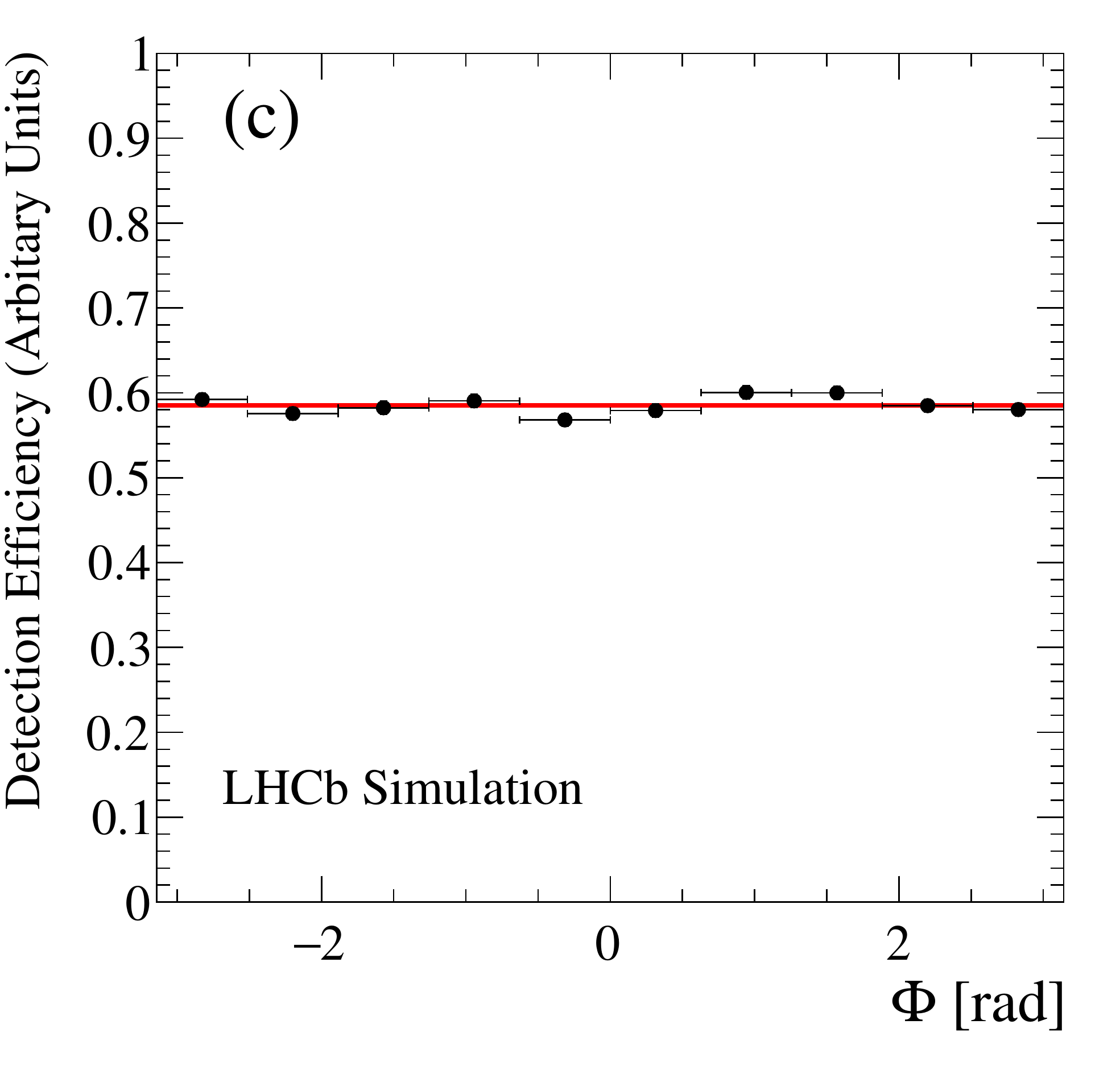} \hfill
\includegraphics[width=0.45\textwidth]{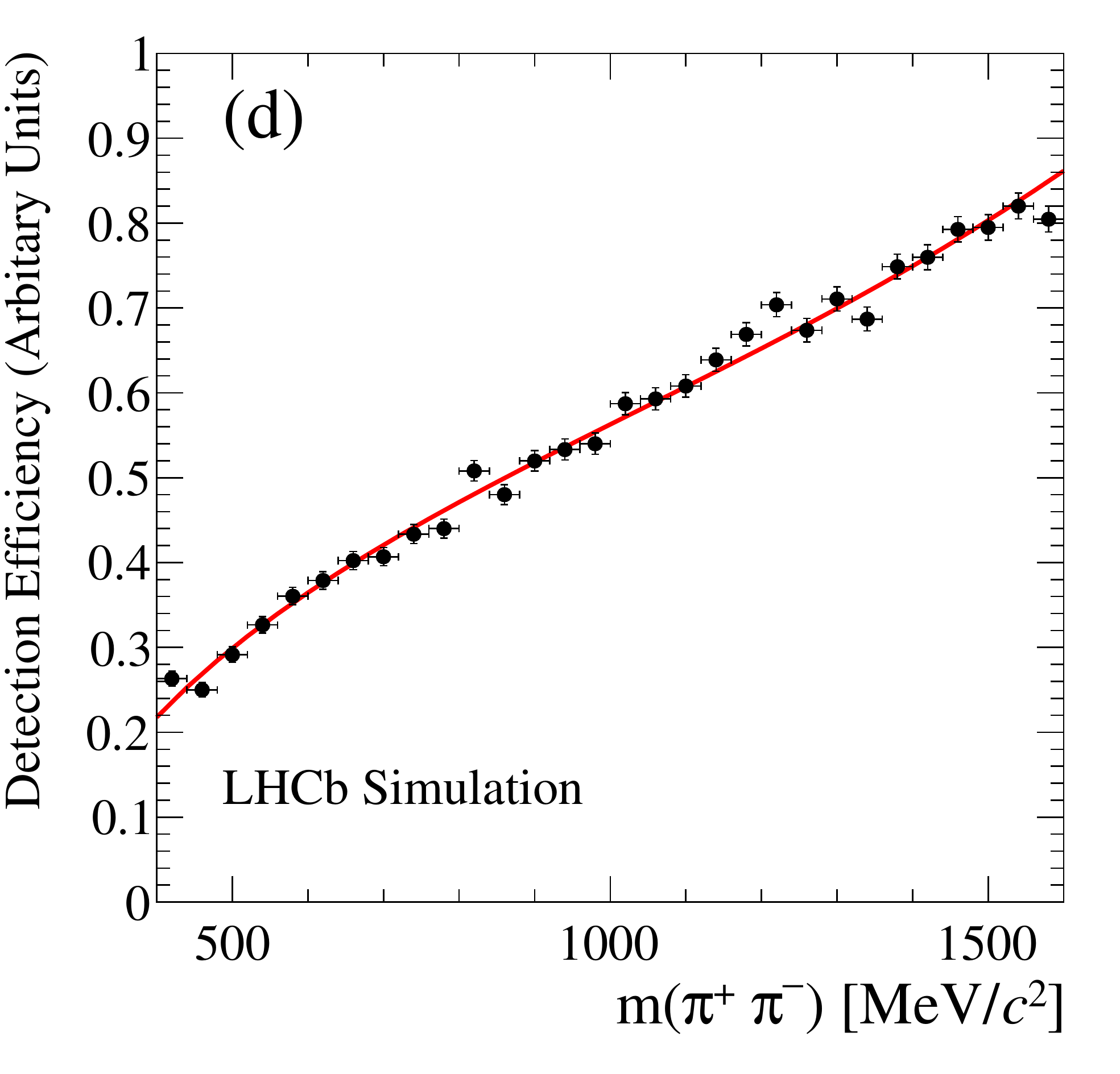}
\caption{\small One-dimensional projections of the detection efficiency 
parameterised using Legendre polynomials (solid red lines) as a function of 
(a) $\cos\theta_1$, (b) $\cos\theta_2$, (c) $\Phi$ 
and (d) $m(\pipi)$, superimposed on the efficiency determined from the 
ratio of the accepted/generated $\BsPhipipi$ events.} 
\label{fig:angularacceptance}
\end{figure}

The LHCb detector geometry and the kinematic selections on the final state particles lead to detection 
efficiencies that vary as a function of $m(\pipi)$ and the decay angles. 
This is studied using simulated signal events, and is parameterised by a four-dimensional function 
using Legendre polynomials, taking into account the correlations 
between the variables. Figure~\ref{fig:angularacceptance} shows the projections of the 
detection efficiency and the function used to describe it. There is a significant drop of efficiency 
at $\cos\theta_1=\pm 1$, a smaller reduction of efficiency for $\cos\theta_2=\pm 1$, 
a flat efficiency in $\Phi$, and a monotonic efficiency increase with $m(\pipi)$. 
This efficiency dependence is included in the amplitude fits.  

The decay rate for the mass range $m(\pipi)<1100$\mevcc can be described primarily by 
the S-wave and P-wave $\pipi$ contributions from the $f_0(980)$ and $\rho$ mesons. 
The S-wave contribution is parameterised by a single amplitude $A_S$.
For the P-wave there are three separate amplitudes $A_0$, $A_\perp$ and $A_{\parallel}$ 
from the possible spin configurations of the final state vector mesons. 
The amplitudes $A_j$, where $j=(0,\perp, \parallel, S)$, are complex and 
can be written as $|A_j|e^{i\delta_j}$. By convention, the phase $\delta_S$ is chosen to be zero.
In the region $m(\pipi)>1100$\mevcc the differential decay rate requires 
additional contributions from the D-wave $f_2(1270)$ meson and other possible 
resonances at higher mass.

The total differential decay rate is given by the square of the sum of the amplitudes. It can be 
written as
\begin{eqnarray}
\frac{d^4\Gamma}{d\cos\theta_1 d\cos\theta_2 d\Phi dm_{\pi\pi}}  =  
\frac{9}{8\pi} \sum_i T_i~f_i( \theta_{1}, \theta_{2}, \Phi )~\mathcal{M}_i(m_{\pi\pi})d\Omega_4(KK\pi\pi)
\hspace{0.1cm} , \label{eqn:decay_sum}
\end{eqnarray}
where the $T_i$ are either squares of the amplitudes $A_j$ or interference terms between them, 
$f_i$ are decay angle distributions, $\mathcal{M}_i$ are resonant $\pipi$ mass distributions and 
$d\Omega_4$ is the phase-space element for four-body decays.    
The detailed forms of these functions are given in Table~2 for the contributions from the 
$f_0(980)$, $\rho$ and $f_2(1270)$ resonances. Note that interference terms between 
\CP-even amplitudes ($A_0$, $A_{\parallel}$, $A_{\perp}^{1270}$) and \CP-odd 
amplitudes ($A_S$, $A_{\perp}$, $A_0^{1270}$, $A_{\parallel}^{1270}$), 
can be ignored in the sum of \Bs and \Bsb decays in the absence of \CP violation, 
as indicated by the measurements in the related decay \BsPhiPhi~\cite{LHCb-PAPER-2014-026}.
With this assumption one \CP-even phase $\delta_{\perp}^{1270}$ can also be chosen to be zero.
The fit neglects the interference terms between P and D-waves, and the P-wave-only interference term 
($i=4$ in Table~2), which are all found to be small when included in the fit. 
This leaves only a single P-wave phase $\delta_{\perp}$ and two D-wave phases $\delta_{\parallel}^{1270}$
and $\delta_0^{1270}$ to be fitted for these three resonant contributions.

\renewcommand{\arraystretch}{1.5}
\begin{table}[tb]
\begin{center}
\caption{\small The individual terms $i=1$ to $i=6$ come from the S-wave and P-wave $\pipi$ 
amplitudes associated with the 
$f_0(980)$ and $\rho$, and the terms $i=7$ to $i=12$ come from the D-wave amplitudes 
associated with the $f_2(1270)$. 
See the text for definitions of $T_i$, $f_i$ and $\mathcal{M}_i$, and for a discussion 
of the interference terms omitted from this table.}
\resizebox{\columnwidth}{!}{
\begin{tabular}{cccc} 
$i$ & $T_i$                 & $f_i$~($\theta_1, \theta_2, \Phi$) & $\mathcal{M}_i(m_{\pi\pi})$ \\ \hline 
1 & $|A_0|^2$         & $\cos^2\theta_1 \cos^2\theta_2$                          & $|M_1(m_{\pi\pi})|^2$ \\
2 & $|A_{\parallel}|^2$ & $\frac{1}{4}\sin^2\theta_1\sin^2\theta_2(1+\cos2\Phi)$ & $|M_1(m_{\pi\pi})|^2$ \\
3 & $|A_\perp|^2$     & $\frac{1}{4}\sin^2\theta_1\sin^2\theta_2(1-\cos2\Phi)$ & $|M_1(m_{\pi\pi})|^2$ \\
4 & $|A_{\parallel}A_0^*|$        & $\sqrt{2}\cos\theta_1\sin\theta_1\cos\theta_2\sin\theta_2\cos\Phi$  & 
$|M_1(m_{\pi\pi})|^2\cos(\delta_{\parallel}-\delta_0)$ \\
5 & $|A_S|^2$           & $\frac{1}{3} \cos^2\theta_2$                                        & $|M_0(m_{\pi\pi})|^2$ \\
6 & $|A_{\perp}A_S^*|$  & $\frac{\sqrt{6}}{3}
\sin\theta_1\cos\theta_2\sin\theta_2\sin\Phi$   & $\Real[M_1(m_{\pi\pi})M_0^*(m_{\pi\pi})e^{i\delta_{\perp}}]$ \\
7& $|A_0^{1270}|^2$    & $\frac{5}{12}(3\cos^2\theta_1-1)^2\cos^2\theta_2$                          & $|M_2(m_{\pi\pi})|^2$ \\
8 & $|A_{\parallel}^{1270}|^2$ & $\frac{5}{2}\sin^2\theta_1\sin^2\theta_2\cos^2\theta_1\cos^2\Phi$ & $|M_2(m_{\pi\pi})|^2$ \\
9 & $|A_\perp^{1270}|^2$     & $\frac{5}{2}\sin^2\theta_1\sin^2\theta_2cos^2\theta_1\sin^2\Phi$ & $|M_2(m_{\pi\pi})|^2$ \\
10 & $|A_{\parallel}^{1270}A_0^{1270*}|$   & $\frac{5}{4\sqrt{6}}(3\cos^2\theta_1-1)\sin 2\theta_1\sin 2\theta_2\cos\Phi$  & 
$|M_2(m_{\pi\pi})|^2\cos(\delta_{\parallel}^{1270}-\delta_0^{1270})$ \\
11 & $|A_{\parallel}^{1270}A_S^*|$  & $\frac{\sqrt{10}}{3}\sin\theta_1\cos\theta_1\sin\theta_2\cos\theta_2\cos\Phi$  & 
$\Real[M_2(m_{\pi\pi})M_0^*(m_{\pi\pi})e^{i\delta_{\parallel}^{1270}}]$ \\
12 & $|A_0^{1270}A_S^*|$ & $\frac{\sqrt{5}}{3}(3\cos^2\theta_1-1)\cos^2\theta_2$ & 
$\Real[M_2(m_{\pi\pi})M_0^*(m_{\pi\pi})e^{-i\delta_0^{1270}}]$ \\ \hline
\end{tabular}
}
\end{center}
\label{tab:decay_rate_list}
\end{table}
\renewcommand{\arraystretch}{1}

Several amplitude fits have been performed including different resonant contributions. 
All fits include the $f_0(980)$ and $f_2(1270)$ resonances. 
The high-mass region $1350 < m(\pipi) < 1600$\mevcc has been modelled by either an S-wave 
or a D-wave $\pipi$ contribution, where the masses and widths of these contributions are determined by the fits, 
but the shapes are constrained to be Breit-Wigner functions. 
In each case the respective terms in Table~2 from $f_0(980)$ or $f_2(1270)$ have to be duplicated
for the higher resonance. For the higher S-wave contribution 
this introduces one new amplitude $A_S^{1500}$ and phase $\delta_S^{1500}$, 
and there is an additional interference term between the two S-wave resonances. 
For the higher D-wave contribution $f_2(1430)$ there are three new amplitudes and phases,
and several interference terms between the two D-wave resonances.
A contribution from the P-wave $\rho(1450)$ has also been considered, 
but is found to be negligible and is not included in the final fit. 
The fit quality has been assessed using a binned $\chi^2$ calculation based 
on the projected $\cos\theta_1$, $\cos\theta_2$ and $m(\pipi)$ distributions. 
In the high-mass region the best fit uses an S-wave component 
with a fitted mass and width of $1427\pm 7$\mevcc and $143\pm 17$\mevcc, hereafter referred 
to as the $f_0(1500)$ for convenience.
The mass is lower than the accepted value of $1504\pm 6$\mevcc for the $f_0(1500)$\cite{PDG2014}. 
It is also lower than the equivalent S-wave component in $\Bs\to\jpsi\pipi$
where the fitted mass and width were $1461\pm 3$\mevcc and $124\pm 7$\mevcc~\cite{LHCb-PAPER-2013-069}. 
This may be due to the absence of contributions from the $\rho$ and $f_2(1270)$ in $\Bs\to\jpsi\pipi$. 
It has been suggested~\cite{Bugg:2008eb, Close:2015rza} that the observed $m(\pipi)$ distributions 
can be described by an interference between the $f_0(1370)$ and $f_0(1500)$, but with the current 
statistics of the $\BsPhipipi$ sample it is not possible to verify this.

\begin{figure}[tb]
\centering
\includegraphics[width=0.45\linewidth]{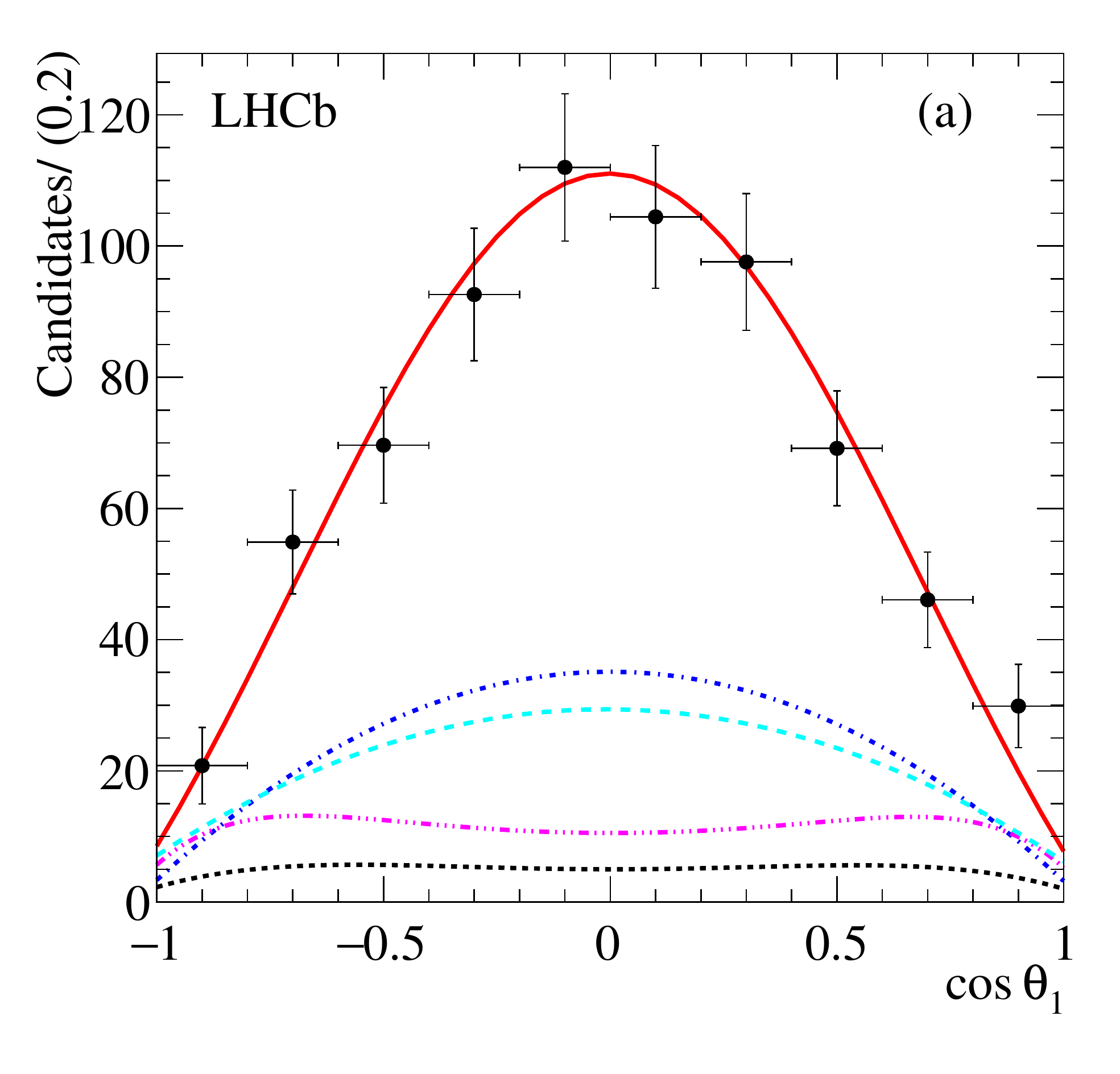} \hskip1cm
\includegraphics[width=0.45\linewidth]{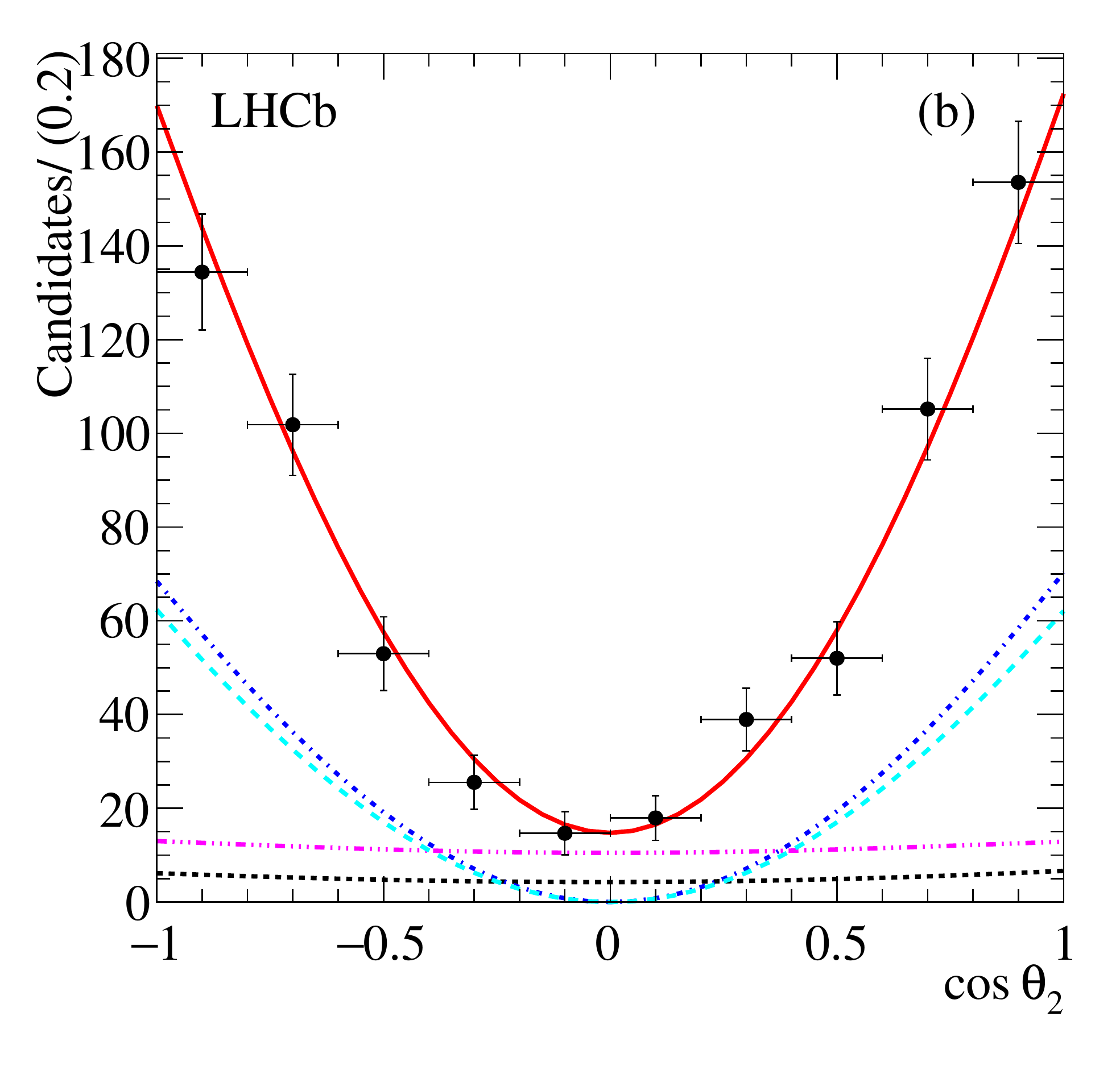} \\
\includegraphics[width=0.45\linewidth]{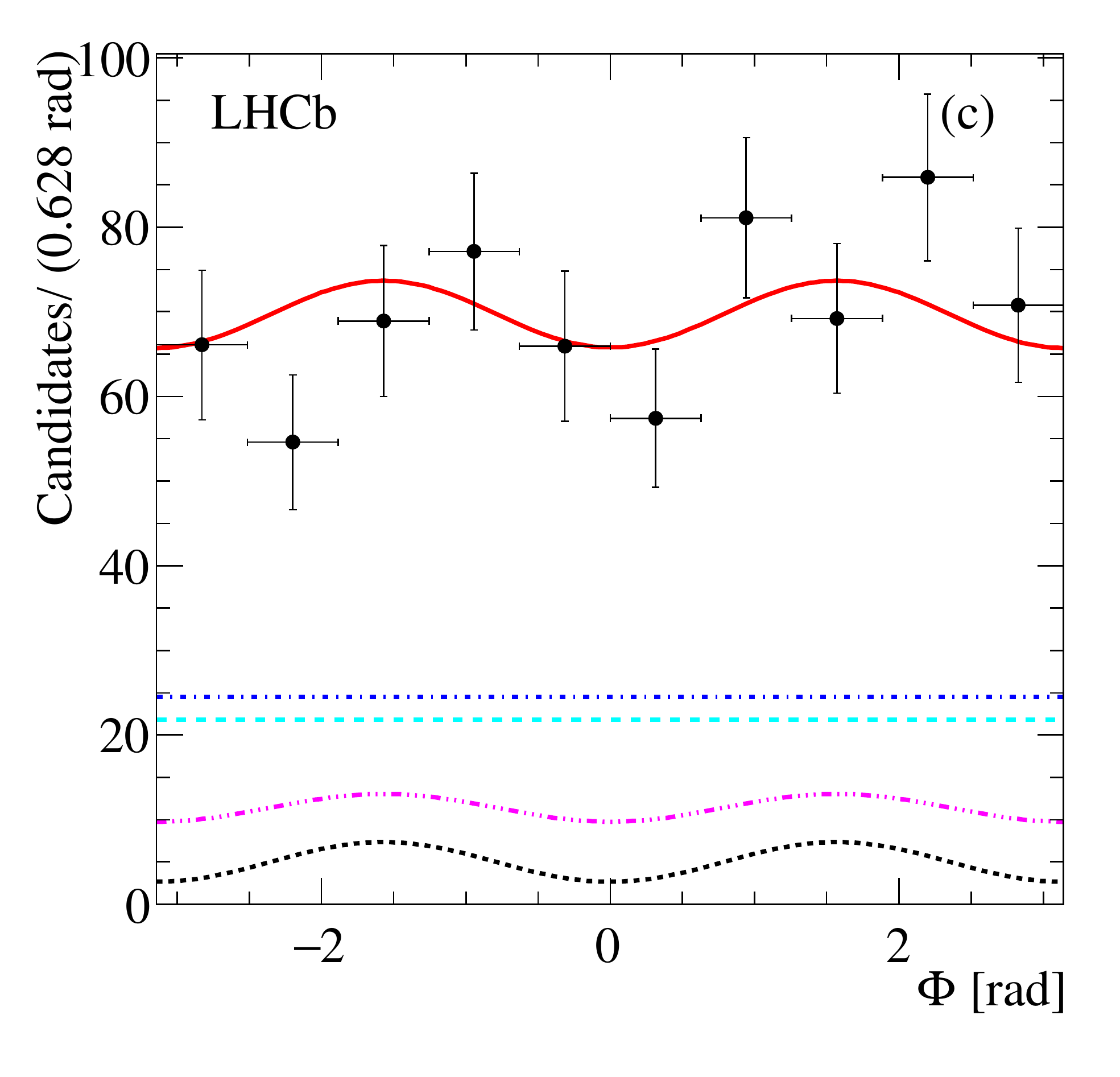}\hskip1cm
\includegraphics[width=0.45\linewidth]{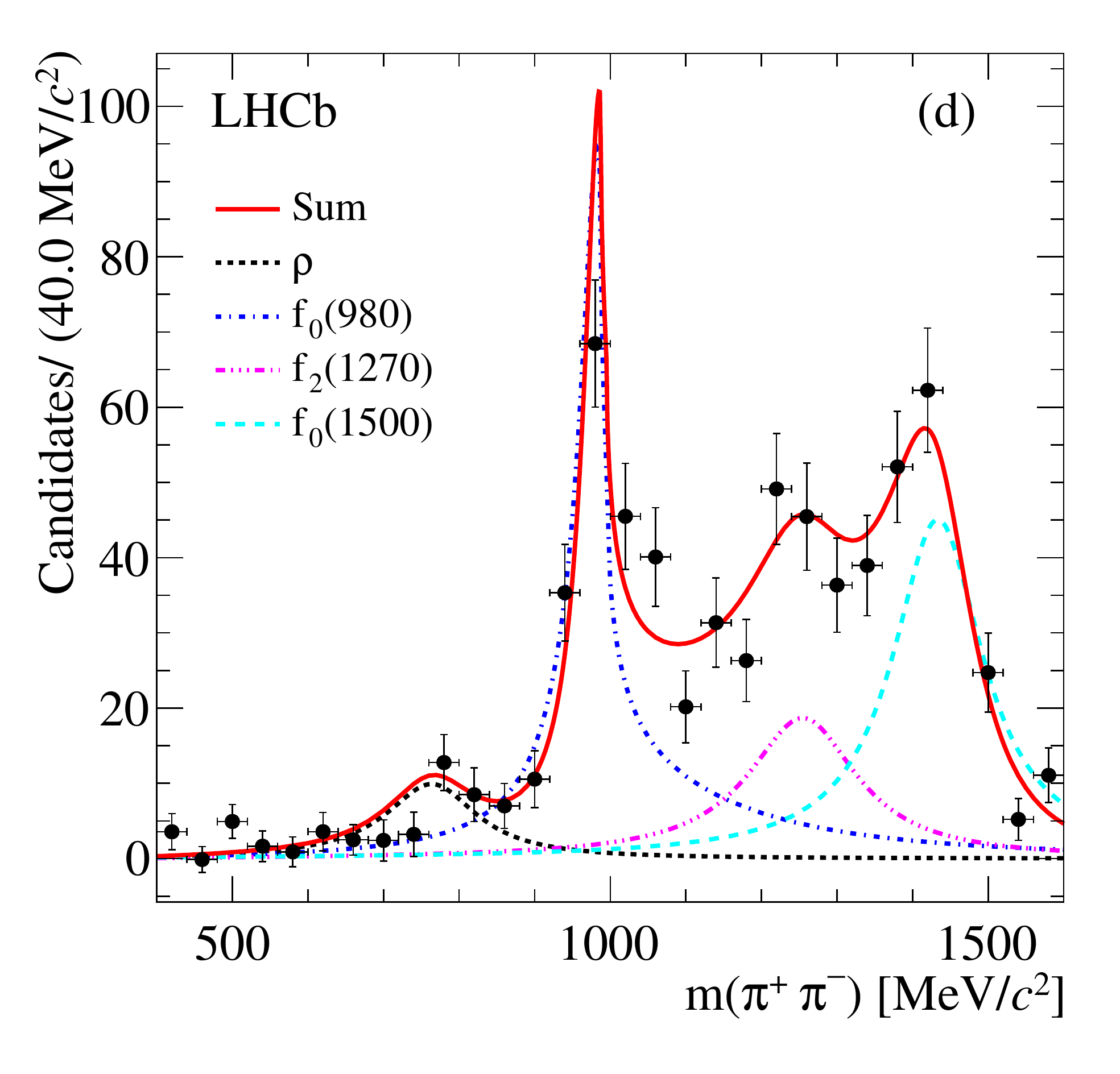}\\
\caption{\small Projections of (a) $\cos\theta_1$, (b) $\cos\theta_2$, (c) $\Phi$, 
and (d) $m(\pipi)$ for the preferred fit. 
The $\rho$ contribution is shown by the dotted (black) line, the $f_0(980)$ by the dot-dashed (blue) line, 
the $f_2(1270)$ by the double-dot-dashed (magenta) line and the $f_0(1500)$ by the dashed (cyan) line. 
Note that the expected distributions from each resonance include the effect of the experimental efficiency. 
The solid (red) line shows the total fit. The points with error bars are the data,  where the background has
been subtracted using the \Bs signal weights from the $\kkpipi $ invariant mass fit.}
\label{fig:angularfit}
\end{figure}

In the low-mass region $m(\pipi)<900$\mevcc the effect of adding a contribution from the 
$\rho$ is studied. The $\rho$ contribution significantly improves the fit quality and 
has a statistical significance of $4.5\sigma$, estimated by running pseudo-experiments. 
A contribution from the $f_0(500)$ has been considered as part of the systematics.
The preferred fit, including the $\rho$, $f_0(980)$, $f_2(1270)$  and $f_0(1500)$,  
has $\chisqndf = 34/20$. Removing the $\rho$ increases 
this to $\chisqndf = 53/24$, and replacing the S-wave $f_0(1500)$ with a 
D-wave $f_2(1430)$ increases it to $\chisqndf = 78/16$.
The projections of the preferred fit, 
including the $\rho$, $f_0(980)$, $f_2(1270)$  and $f_0(1500)$, 
are shown in Fig.~\ref{fig:angularfit}. 
The fitted amplitudes and phases are given in Table~\ref{tab:fitamplitudes}.
From Fig.~\ref{fig:angularfit} it can be seen that the low numbers of observed candidates
in the regions $|\cos\theta_1|>0.8$ and $|\cos\theta_2|<0.4$ require a large 
S-wave $\pipi$ contribution, and smaller P-wave and D-wave contributions. 

\begin{table}[tb]
\begin{center}
\caption{\small The resonance amplitudes and phases from the preferred fit
to the $m(\pipi)$ and decay angle distributions of the \Bs candidates, 
including the $\rho$, $f_0(980)$, $f_2(1270)$  and $f_0(1500)$.
See text for definitions of the amplitudes and phases.}
\begin{tabular}{cc|cc}
Amplitude & Fit value &  Phase & Fit value (rad)\\
\hline 
$A_0$  &  $0.212\pm 0.035$  &  & \\
$A_{\parallel}$ & $0.049\pm 0.031$ & & \\
$A_{\perp}$ &  $0.168\pm 0.026$ & $\delta_{\perp}$ &  $+1.90\pm 0.28$ \\  
$A_S$  &  $0.603\pm 0.036$  & &  \\    
$A_0^{1270}$ & $0.295\pm 0.058$ & $\delta_0^{1270}$ &  $-0.62\pm 0.18$ \\
$A_{\parallel}^{1270}$ & $0.203\pm 0.042$ & $\delta_{\parallel}^{1270}$ &  $+1.26\pm 0.25$ \\
$A_{\perp}^{1270}$ & $0.261\pm 0.037$ & & \\
$A_S^{1500}$  &  $0.604\pm 0.031$  & $\delta_S^{1500}$ & $+3.14\pm 0.30$ \\ \hline
\end{tabular}
\label{tab:fitamplitudes}
\end{center}
\end{table}

To convert the fitted amplitudes into fractional contributions from different resonances 
they need to be first summed over the different polarisations and then squared. 
Interference terms between the resonances are small, but not completely negligible. 
When calculating the fit fractions and event yields, the interference terms are included 
in the total yield but not in the individual resonance yields. As a consequence, the sum 
of the fractions is not 100\%. Table~\ref{tab:fitfractions} gives the fit fractions 
and the corresponding event yields for the resonant contributions to the $\BsPhipipi$ decay 
for the fits with and without a $\rho$. 

\begin{table}[tb]
\centering
\caption{Fit fractions in \% and event yields for the resonances contributing to $\BsPhipipi$.
Results are quoted for the preferred model with a $\rho$, and for an alternative model without 
a $\rho$ which is used to evaluate systematic uncertainties.}
\begin{tabular}{l|cc|cc} 
Resonance & \multicolumn{2}{c|}{Fit fractions \%} &  \multicolumn{2}{c}{Event yields} \\ 
contribution & without $\rho$ & with $\rho$ & without $\rho$ & with $\rho$ \\ \hline
$\rho$   &	 --	    &  7.1 $\pm$ 1.5 &  --        &  50 $\pm$ 11 \\
$f_0(980)$ & 39.5 $\pm$ 2.9 & 35.6 $\pm$ 4.3 & 274 $\pm$ 23 & 247 $\pm$ 31 \\
$f_2(1270)$ & 23.5 $\pm$ 2.7 & 15.1 $\pm$ 3.2 & 163 $\pm$ 20 & 112 $\pm$ 23\\
$f_0(1500)$ & 26.5 $\pm$ 2.2 & 34.7 $\pm$ 3.4 & 184 $\pm$ 17 & 241 $\pm$ 26 \\ \hline
\end{tabular}
\label{tab:fitfractions}
\end{table}

%% file: results.tex
\section{Determination of branching fractions}

\label{sec:Results}

The  branching fractions are determined using the relationship
\begin{displaymath}
\frac{\mathcal{B}(\Bs(\Bz)\to\phi\pipi)}{\mathcal{B}(B_s^0 \to \phi \phi)}
= \frac{N(\phi \pipi)}{N(\phi \phi)} \times \frac{\varepsilon^{tot}_{\phi \phi}}{\varepsilon^{tot}_{\phi \pipi}} 
\times  \frac{f_s}{f_d} \times \mathcal{B}(\phi \to K^+ K^-) \times f_P \hspace{0.1cm} .
\end{displaymath}
The signal yields $N(\phi \pipi)$ for the inclusive modes are taken from the fit to the \kkpipi mass distribution
in Fig.~\ref{fig:massfit2}, and for the normalisation mode $N(\phi \phi)$ is taken from the fit to 
the $K^+K^-K^+K^-$ mass distribution in Fig.~\ref{fig:phiphi_mass}.
The factor $f_P = (93\pm 4)$\% corrects for the difference in the fitted 
S-wave \kk contributions to the $K^+K^-$ mass distribution around the nominal $\phi$ mass
between the signal and normalisation modes. 
The branching fraction $\mathcal{B}(\phi \to K^+ K^-) = (48.9 \pm 0.5)\%$\cite{PDG2014} enters twice in 
the normalisation mode. The factor $f_s/f_d=0.259\pm 0.015$\cite{LHCb-PAPER-2012-037} 
only applies to the $\BdPhipipi$ mode in the above ratio, but also appears in the ratio of 
$\BsPhiPhi$ relative to $\Bz\to\phi\Kstar$, so it effectively cancels out in the determination 
of the $\BdPhipipi$ branching fraction. For the $\BsPhipipi$ mode it is included in the determination 
of $\mathcal{B}(\BsPhiPhi)$~\cite{LHCb-PAPER-2015-028}. 
The total selection efficiencies $\varepsilon^{{\rm tot}}_{\phi \pipi}$ and 
$\varepsilon^{{\rm tot}}_{\phi \phi}$ are given in Table~\ref{tab:sel_eff}. 

For the inclusive modes the branching fractions with $400<m(\pipi)<1600\mevcc$ are
\begin{displaymath}
{\cal{B}}(\BsPhipipi) = [3.48 \pm 0.23]\times 10^{-6} \hspace{0.1cm} ,
\end{displaymath}
and
\begin{displaymath}
{\cal{B}}(\BdPhipipi) = [1.82 \pm 0.25]\times 10^{-7} \hspace{0.1cm} ,
\end{displaymath}
where the quoted uncertainties are purely statistical, but include the uncertainties on 
the yield of the normalisation mode, and on the S-wave $K^+K^-$ contributions to the signal and 
normalisation modes. For the exclusive \Bs modes the signal yields are taken from the final column 
in Table~\ref{tab:fitfractions}. The branching fractions are
\begin{displaymath}
{\cal{B}}(\BsPhifO) = [1.12 \pm 0.16]\times 10^{-6} \hspace{0.1cm} ,
\end{displaymath}
\begin{displaymath} 
{\cal{B}}(\Bs\to\phi\ftwotw) = [0.61 \pm 0.13]\times 10^{-6} \hspace{0.1cm} ,
\end{displaymath}
and 
\begin{displaymath}
{\cal{B}}(\BsPhiRho^0) = [2.7 \pm 0.6]\times 10^{-7}\hspace{0.1cm} .
\end{displaymath}
The remaining $1.5\times 10^{-6}$ of the inclusive $\Bs$ branching fraction is mostly accounted for 
by an S-wave contribution in the region $1350-1600\mevcc$ as discussed in the previous section. 

\begin{table} [tb]
\begin{center}
\caption{\small Selection efficiencies for the signal and normalisation modes in \%, 
as determined from simulated event samples. Here ``Initial selection" refers to a loose set of requirements 
on the four tracks forming the \B candidate. The ``Offline selection" includes the charm 
and $\phi K^*{^0}$ vetoes, as well as the BDT. 
Angular acceptance and decay time refer to corrections made 
for the incorrect modelling of these distributions in the inclusive and 
$\BsPhifO$ simulated event samples.}    
\begin{tabular}{ccccc}
Efficiency &  $\Bs(\Bz)\to\phi\pipi$ & $B_s^0 \to \phi \rho^0$ & 
\BsPhifO & \BsPhiPhi \\
\hline 
Detector acceptance  &  17.4  &  18.1 &  18.0  &   17.1\\    
Initial selection    &   8.43  &  7.35 &  8.48 &   14.6\\
Trigger              &  34.9  &  34.9 &  34.5 &    28.6\\
Offline selection    &  63.9 (57.1)  &  62.5 &  63.2 &    59.3 \\ 
Particle identification & 87.5  &  87.5 &  87.5 &  93.9 \\ 
Angular acceptance  & 95.9 (100) & 100 & 100 & 100 \\
Decay time     & 100 & 100 & 104.5  & 100 \\ \hline
Total                &  0.275 (0.256) & 0.254 &  0.303 &   0.398 \\ \hline
\end{tabular}
\label{tab:sel_eff}
\end{center}
\end{table}

%% file: systematics.tex
\section{Systematic uncertainties}

\label{sec:Systematics}

Many systematic effects cancel in the ratio of efficiencies between the signal
and normalisation modes. The remaining systematic uncertainties in the determination of 
the branching fractions come from replacing the $\pipi$ 
pair with a second $\phi$ meson decaying to two kaons.  
The systematic uncertainties are summarised in Table~\ref{tab:systematics}.  

The trigger selection has a different performance for the $\Bs\to\phi\pipi$ signal
and for the $\Bs\to\phi\phi$ normalisation mode due to the different kinematics of 
the final state hadrons. The simulation of the trigger does not reproduce 
this difference accurately for hadronic decays, 
and a $\Dz\to K^-\pi^+$ control sample, collected with a minimum bias trigger, 
is used to evaluate corrections to the trigger efficiencies between the simulation and the data. 
These are applied as per-event reweightings of the simulation as a function of track $p_T$, 
particle type $K$ or $\pi$, and magnetic field orientation.
For both the signal and normalisation modes there are large corrections of $\approx 30$\%, 
but they almost completely cancel in the ratio, leaving 
a systematic uncertainty of 0.5\% from this source.  

Another aspect of the detector efficiency that is not accurately modelled by the simulation is 
hadronic interactions in the detector. A sample of simulated
$\Bz\to \jpsi\Kstar^0$ events is used to determine the fraction 
of kaons and pions that interact within the detector as a function of their momentum.
On average this varies from 11\% for $K^+$ to 15\% for $\pi^-$. 
These numbers are then scaled up to account for 
additional material in the detector compared to the simulation. 
The effect partly cancels in the ratio  
of the signal and normalisation modes leaving a 0.5\% systematic uncertainty from this source. 

The offline selection efficiency has an uncertainty coming from the performance of 
the multivariate BDT. This has been studied by varying the selection on the 
\BsPhiPhi normalisation mode,  
and extracting the shapes of the input variables from data using the \sPlot technique.
The distributions agree quite well between simulation and data, but there are 
small differences. When these are propagated to the signal modes they lead to  
a reduction in the BDT efficiency. Again the effect partially cancels in the ratio  
leaving a systematic uncertainty of 2.3\%. 

The offline selection also has an uncertainty coming from the different particle identification
criteria used for the $\pipi$ in the signal and the $K^+K^-$ from the 
second $\phi$ in the normalisation mode. 
Corrections between simulation and data are studied using calibration samples, 
with kaons and pions binned in $p_T$, $\eta$ and number of tracks in the event.
There is an uncertainty of 0.1\% from the size of the calibration samples.
Using different binning schemes for the corrections leads to a slightly higher 
estimate for the systematic uncertainty of 0.3\%. 

For the angular acceptance there is an uncertainty in the $m(\pipi)$ and 
angular distributions for the inclusive decays, and in the polarisations of 
the $\rho^0$ and $f_2(1270)$. A three-dimensional binning in 
$[\cos\theta_1,\cos\theta_2, m(\pipi)]$ is used to reweight the 
simulation to match the data distributions for these modes.
The accuracy of this procedure is limited by the number of bins and hence by the data statistics.
By varying the binning scheme systematic uncertainties of 3.8\% (10.7\%) are determined 
for $\Bs$ ($\Bd$) from this reweighting procedure. The larger $\Bd$ uncertainty reflects 
the smaller signal yield. The angular distribution of the $\Bs\to\phi\phi$ normalisation mode
is modelled according to the published LHCb measurements~\cite{LHCb-PAPER-2014-026}, 
which introduces a negligible uncertainty.   

The decay time acceptance of the detector falls off rapidly at short decay times
due to the requirement that the tracks are consistent with coming from a secondary vertex. 
For \Bs decays the decay time distribution is modelled by the flavour-specific lifetime, 
but it should be modelled by a combination of the heavy and light mass eigenstates, 
depending on the decay mode. A systematic uncertainty of 1.1\% is found when replacing the 
flavour-specific lifetime by the lifetime of the heavy eigenstate and determining the change 
in the decay time acceptance. There is no effect on \Bd decays or on the normalisation mode 
where the lifetime is modelled according to the published measurements.  

The $K^+K^-\pipi$  and $K^+K^-K^+K^-$ invariant mass fits are repeated using a 
single Gaussian and using a power-law function to model the tails of the signal shapes.
For the $m(K^+K^-\pipi)$ fit contributions from
partially reconstructed backgrounds are added, including $\Bs\to\phi\phi(\pi^+\pi^-\pi^0)$ and 
$\Bs\to\phi\eta'(\pi^+\pi^-\gamma)$.
% and the fit is done over a wider mass range with $m(K^+K^-\pipi)<5100$\mevcc. 
These changes lead to uncertainties 
on the $\Bs$ ($\Bd$) yields of 1.2\% (19.5\%). The large uncertainty on the \Bd yield 
comes both from the change in the signal shape and from the addition of partially 
reconstructed \Bs backgrounds. This systematic uncertainty reduces the significance of 
the \Bd signal from $7.7\sigma$ to $4.5\sigma$. 

The results of the amplitude analysis for the exclusive $\Bs$ decays depend on the 
set of input resonances that are used. The effect of including the $\rho^0$ is treated 
as a systematic uncertainty on the $f_0(980)$ and $f_2(1270)$ yields (see Table~\ref{tab:fitfractions}). 
The effect of adding either an $f_0(500)$ or a $\rho(1450)$ 
is treated as a systematic uncertainty on all the exclusive modes. 
       
The difference between the S-wave $K^+K^-$ components in the signal and normalisation modes 
is measured to be $(7.1\pm 4.0)$\% from fits to the $K^+K^-$ mass distributions.
The uncertainty on this is treated as part of the statistical error. 
However, the S-wave component of the signal sample was not included 
in the amplitude analysis where it would give a flat distribution in $\cos\theta_2$. 
A study of the dependence of the S-wave $K^+K^-$ component as a function of $m(\pipi)$
does not indicate a significant variation, and the statistical uncertainty of 6\%  from this 
study is taken as a systematic uncertainty on the yields of the exclusive modes extracted from the 
amplitude analysis.

\begin{table} [tb]
\begin{center}
\caption{\small Systematic uncertainties in \% on the branching fractions of \Bs and \Bz decays.
All the uncertainties are taken on the ratio of the signal to the normalisation mode.
Uncertainties marked by a dash are either negligible or exactly zero. 
The asymmetric uncertainties on $\phi f_0(980)$ and $\phi f_2(1270)$ come from the differences in 
yields between the fits with and without the $\rho^0$ contribution.}
\begin{tabular}{ccccc}
Systematic &  $B_s^0 \to \phi \rho^0$ & $\BsPhifO$ &  $B_s^0 \to \phi \ftwotw$ & 
$\Bs(\Bz)\to\phi\pipi$ \\ \hline
Trigger &  0.5  &  0.5 &  0.5 & 0.5  \\    
Hadronic interactions & 0.5 & 0.5 & 0.5 & 0.5\\
Offline selection &  2.3 & 2.3 & 2.3 & 2.3 \\
Particle identification & 0.3 & 0.3 & 0.3 & 0.3 \\
Angular acceptance & 3.8 &  $-$ & 3.8 & 3.8 (10.7) \\ 
Decay time acceptance & 1.1 & 1.1 & 1.1 &  1.1 ($-$)\\
$m(K^+K^-\pipi)$ fit & 1.2 & 1.2 & 1.2 & 1.2 (19.5) \\
Amplitude analysis & 2.5 & $+4.7/-0.4$ & $+17.6/-2.7$  & $-$ \\
S-wave $K^+K^-$ & 6.0 & 6.0 & 6.0 & $-$ \\ \hline
Total    & 7.0 & $+8.2/-6.7$ & $+19.2/-8.1$ & 4.8 (22.4) \\ \hline
\end{tabular}  
\label{tab:systematics}
\end{center}
\end{table}

%% file: conclusions.tex
\section{Summary and conclusions}
\label{sec:conclusions}
This paper reports the first observation of the 
inclusive decay $\BsPhipipi$. 
The branching fraction in the mass range $400< m(\pipi) <1600$\mevcc
is measured to be
\begin{displaymath}
\mathcal{B}(\BsPhipipi) = [3.48 \pm 0.23 \pm 0.17 \pm 0.35] \times 10^{-6} \; ,
\end{displaymath}
where the first uncertainty is statistical, the second  is systematic, and the third 
is due to the  normalisation mode $\BsPhiPhi$.
  
Evidence is also seen for the inclusive decay $\BdPhipipi$ with a statistical 
significance of 7.7$\sigma$, which is reduced to 4.5$\sigma$ after taking into account 
the systematic uncertainties on the signal yield. 
The branching fraction in the mass range $400< m(\pipi) <1600$\mevcc is
\begin{displaymath}
\mathcal{B}(\BdPhipipi) = [1.82 \pm 0.25 \pm 0.41 \pm 0.14] \times 10^{-7} \; .
\end{displaymath}
An amplitude analysis is used to separate out exclusive contributions to the \Bs decays. 
The decay \BsPhifO is observed with a significance of $8\,\sigma$,
and the product branching fraction is
\begin{displaymath}
\mathcal{B}(\BsPhifO, \fO \to \pipi) = [1.12 \pm 0.16 ^{+0.09}_{-0.08} \pm 0.11] \times 10^{-6} \; .
\end{displaymath}
The decay $B_s^0 \to \phi \ftwotw$  is observed with a significance of
$5\,\sigma$, and the product branching fraction is
\begin{displaymath}
\mathcal{B}(B_s^0 \to \phi \ftwotw, \ftwotw \to \pipi) = [0.61 \pm 0.13 ^{+0.12}_{-0.05} \pm 0.06] \times 10^{-6} \, .
\end{displaymath}
There is also a contribution from higher mass S-wave $\pipi$ states in the region $1350-1600\mevcc$, 
which could be ascribed to a linear superposition of the $f_0(1370)$ and the $f_0(1500)$.  
There is $4\sigma$ evidence for the decay $\BsPhiRho^0$ with a branching fraction of
\begin{displaymath}
\mathcal{B}(\BsPhiRho^0) = [2.7 \pm 0.7 \pm 0.2 \pm 0.2] \times 10^{-7} \, .
\end{displaymath}
This is lower than the Standard Model prediction of $[4.4^{+2.2}_{-0.7}]\times 10^{-7}$,
but still consistent with it, 
and provides a constraint on possible contributions from new physics in this decay. 

With more data coming from the LHC it will be possible to further investigate the exclusive 
decays, perform an amplitude analysis of the \Bz decays, 
and eventually make measurements of time-dependent \CP violation that are 
complementary to the measurements already made in the $\BsPhiPhi$ decay.

%% file: acknowledgements.tex
\section{Acknowledgements}

\noindent We express our gratitude to our colleagues in the CERN
accelerator departments for the excellent performance of the LHC. We
thank the technical and administrative staff at the LHCb
institutes. We acknowledge support from CERN and from the national
agencies: CAPES, CNPq, FAPERJ and FINEP (Brazil); NSFC (China);
CNRS/IN2P3 (France); BMBF, DFG and MPG (Germany); INFN (Italy); 
FOM and NWO (The Netherlands); MNiSW and NCN (Poland); MEN/IFA (Romania); 
MinES and FASO (Russia); MinECo (Spain); SNSF and SER (Switzerland); 
NASU (Ukraine); STFC (United Kingdom); NSF (USA).
We acknowledge the computing resources that are provided by CERN, IN2P3 (France), 
KIT and DESY (Germany), INFN (Italy), SURF (The Netherlands), PIC (Spain), GridPP (United Kingdom), RRCKI and Yandex LLC (Russia), CSCS (Switzerland), IFIN-HH (Romania), CBPF (Brazil), PL-GRID (Poland) and OSC (USA). We are indebted to the communities behind the multiple open 
source software packages on which we depend.
Individual groups or members have received support from AvH Foundation (Germany),
EPLANET, Marie Sk\l{}odowska-Curie Actions and ERC (European Union), 
Conseil G\'{e}n\'{e}ral de Haute-Savoie, Labex ENIGMASS and OCEVU, 
R\'{e}gion Auvergne (France), RFBR and Yandex LLC (Russia), GVA, XuntaGal and 
GENCAT (Spain), Herchel Smith Fund, The Royal Society, 
Royal Commission for the Exhibition of 1851 and the Leverhulme Trust (United Kingdom).

%% file: LHCb_Authorship_flat_29-Jun-2016.tex
\centerline{\large\bf LHCb collaboration}
\begin{flushleft}
\small
R.~Aaij$^{40}$,
B.~Adeva$^{39}$,
M.~Adinolfi$^{48}$,
Z.~Ajaltouni$^{5}$,
S.~Akar$^{6}$,
J.~Albrecht$^{10}$,
F.~Alessio$^{40}$,
M.~Alexander$^{53}$,
S.~Ali$^{43}$,
G.~Alkhazov$^{31}$
% ...

{\footnotesize \it
% ...
$ ^{5}$Clermont Universit{\'e}, Universit{\'e} Blaise Pascal, CNRS/IN2P3, LPC, Clermont-Ferrand, France\\
$ ^{6}$CPPM, Aix-Marseille Universit{\'e}, CNRS/IN2P3, Marseille, France\\
% ...
$ ^{10}$Fakult{\"a}t Physik, Technische Universit{\"a}t Dortmund, Dortmund, Germany\\
% ...
$ ^{31}$Petersburg Nuclear Physics Institute (PNPI), Gatchina, Russia\\
% ...
$ ^{39}$Universidad de Santiago de Compostela, Santiago de Compostela, Spain\\
$ ^{40}$European Organization for Nuclear Research (CERN), Geneva, Switzerland\\
% ...
$ ^{43}$Nikhef National Institute for Subatomic Physics, Amsterdam, The Netherlands\\
% ...
$ ^{48}$H.H. Wills Physics Laboratory, University of Bristol, Bristol, United Kingdom\\
% ...
$ ^{53}$School of Physics and Astronomy, University of Glasgow, Glasgow, United Kingdom\\
% ...
}
\end{flushleft}